\documentclass[12pt]{article}

\usepackage{epsfig}

\def \sect #1 {\setcounter{equation} 0\section{#1}}
\def \be  {\begin{equation}}
\def \ee  {\end{equation}}
\def \ba  {\begin{eqnarray}}
\def \ea  {\end{eqnarray}}
\def \baa {\begin{eqnarray*}}
\def \eaa {\end{eqnarray*}}
\def \bb  {\begin{thebibliography}}
\def \eb  {\end{thebibliography}}

\def \lab #1 {\label{#1}}

\def \fracs #1#2 {\mbox{\small $\frac{#1}{#2}$}}

\def \bin #1#2 {{\left({#1}\atop{#2}\right)}}
\def \as {\relax\ifmmode\alpha_s\else{$\alpha_s${ }}\fi}

\def \al #1 {\frac {\as({#1})}{\pi} }
\def \ds #1 {\ooalign{$\hfil/\hfil$\crcr$#1$}}

\def \QCD {\mbox{{\tiny QCD}}}

\newcommand \bit {\begin{itemize}}
\newcommand \eit {\end{itemize}}

\newcommand \bea{\begin{eqnarray}}
\newcommand \eea{\end{eqnarray}}
\newcommand \ok {{\omega}_k}
\newcommand \ol {{\omega}_l}
\newcommand \sfi{\sin \phi}
\newcommand \cfi{\cos \phi}

\newcommand \sd {\sin \delta}

\newcommand \de {\Delta \eta}
\newcommand \xic {{\hat \xi}_c}
\newcommand \Ncol {{\mathcal{N}}_C}

\def \O {\Omega}
\def \o {\omega}

\def\hepph  #1 {{\tt hep-ph/#1}}

\begin{document}

\begin{flushright}
YITP-03-06 \\
\today
\end{flushright}

\vspace*{30mm}
\begin{center}
{\LARGE \bf Event Shape/Energy Flow Correlations} \vspace*{2mm} \\

\par\vspace*{20mm}\par

{\large  \bf
Carola F.\ Berger,
    Tibor K\'ucs, and
George Sterman}

\bigskip

{\em C.N.\ Yang Institute for Theoretical Physics,
Stony Brook University, SUNY\\
Stony Brook, New York 11794 -- 3840, U.S.A.}

\vspace*{8mm}

\begin{abstract}
We introduce a set of correlations between energy flow
and event shapes that are sensitive to the flow
of color at short distances in jet events.  These correlations are
formulated for a general set of
event shapes, which includes jet broadening and thrust as special cases.
We illustrate the method for $e^+e^-$ dijet events, and calculate the
correlation at leading logarithm in the energy flow and
at next-to-leading-logarithm in the event shape.
\end{abstract}

\end{center}

\vspace*{8mm}

\section{Introduction}

The agreement of
theoretical predictions  with experiment for jet cross sections is often
impressive.
This is especially so for inclusive jet cross sections at
high $p_T$, using fixed-order factorized perturbation theory
and parton distribution functions \cite{tevajet}.  A
good deal is also known about the substructure of jets,
through the theoretical and  experimental study
of multiplicity distributions and fragmentation functions \cite{DKTrev},
and of event shapes \cite{eshape,farhi,irsproof}.   Event shape
distributions \cite{thrustresum,broaden1,broaden2} in
particular offer a bridge between the perturbative, short-distance and
the nonperturbative, long-distance dynamics  of QCD \cite{ptnpdist}.

Energy flow \cite{flow} into angular regions between energetic
jets gives information that is in some ways complementary
to what we learn from event shapes.  In perturbation theory,
the distribution of particles in the final state reflects
interference between radiation from different jets \cite{DKTrev},
and there is ample evidence for perturbative
antenna patterns in interjet radiation at both $\rm e^+e^-$ \cite{elexp}
and hadron colliders \cite{EKSWjet,D0}.  Energy flow between
jets must also encode the mechanisms that neutralize color in the
hadronization process, and the transition of QCD from weak
to strong coupling.
Knowledge of the interplay
between energy and color flows \cite{KOS,BKS1} may help
identify the underlying
event in hadron collisions \cite{underlying},
to distinguish QCD bremsstrahlung
from signals of new physics.  Nevertheless, the systematic computation of
energy flow into interjet regions has turned out to be
subtle \cite{DS} for reasons that we will review below,
and requires a careful construction of the class of
jet events.  It is the purpose of this work to provide
such a construction, using event shapes as a tool.

In this paper, we introduce correlations
between event shapes and energy flow, ``shape/flow correlations",
    that are sensitive primarily to radiation from the highest-energy
jets.  So long as
the observed energy is not too small, in a manner to be quantified
below, we may control logarithms of the ratio of energy flow
to jet energy \cite{BKS1,BKSproc}.

The energy flow observables that we discuss
below are distributions
associated with radiation into a chosen interjet angular region,
$\O$.  Within $\O$ we identify a kinematic quantity $Q_\O\equiv
\varepsilon Q$, at c.m.\ energy $Q$, with
$ \varepsilon\ll 1$.  $Q_\O$ may be the sum of energies, transverse
energies or related
observables for the particles
emitted into $\O$.  Let us denote by $\bar\O$ the complement
of $\O$.  We are interested in the distribution of $Q_\O$
for events with a fixed number of jets in $\bar \O$.
This set of events may be represented  schematically as
\begin{equation}
A + B \rightarrow \mbox{  Jets }  + X_{\bar{\O}}
    + R_\O (Q_\O)\, .
\label{event}
\end{equation}
Here $X_{\bar\O}$ stands for radiation into the regions
between $\O$ and the jet axes, and $R_\O$ for
radiation into $\O$.

\begin{figure}[htb]
\begin{center}
\epsfig{file=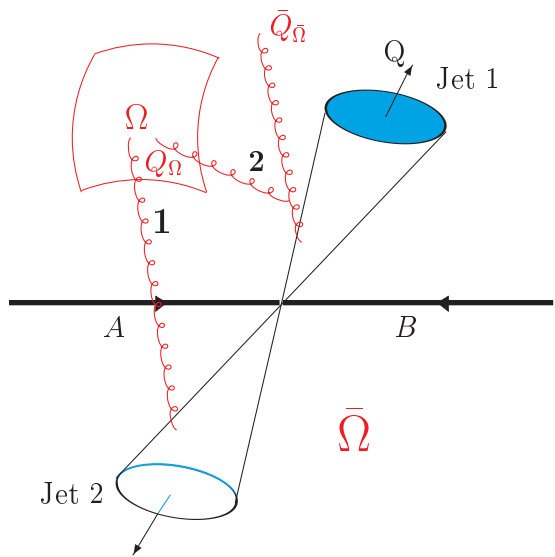,height=8cm,clip=0}
\caption{Sources of global and non-global logarithms in dijet events.
Configuration 1, a
primary emission, is the source of global logarithms.  Configuration 2
can give non-global logarithms.}
\label{eventfig}
\end{center}
\end{figure}

The subtlety associated with the computation of energy flow
concerns the origin of logarithms, and is illustrated by
Fig. \ref{eventfig}.
Gluon 1 in Fig.\ \ref{eventfig} is
an example of a primary gluon,
emitted directly from
the hard partons  near a jet axes.
Phase space integrals for primary emissions contribute single logarithms
per loop: $(1/Q_\O)\as^n \ln^{n-1} (Q/Q_\O) =
(1/\varepsilon Q)\as^n\ln^{n-1}(1/\varepsilon)$, $n\ge 1$, and
these logarithms exponentiate in a straightforward fashion \cite{BKS1}.
At fixed $Q_\O$
for Eq.\ (\ref{event}), however, there is another source of
potentially large logarithmic
corrections in $Q_\O$.  These are illustrated by gluon 2
in the figure, an example of
secondary radiation in $\O$, originating a parton emitted
by one  of the leading jets that define the event into intermediate region
    $\bar{\O}$.
As observed by Dasgupta and
Salam \cite{DS}, emissions into $\O$ from such secondary
partons   can also result in logarithmic corrections, of the form
$(1/Q_\O)\as^n \ln^{n-1}(\bar{Q}_{\bar{\O}}/Q_\O)$, $n\ge 2$,
where $\bar{Q}_{\bar{\O}}$ is the maximum energy
emitted into $\bar{\O}$.  These logarithms arise
from strong ordering in the energies of the primary
and secondary radiation
because real and virtual enhancements
associated with secondary emissions do not
cancel each other fully at fixed $Q_\O$.

If the cross section is
fully inclusive outside of $\O$, so that no restriction
is placed on the radiation into $\bar{\O}$,
$\bar{Q}_{\bar{\O}}$ can approach $Q$, and
the secondary logarithms can become as important as
the primary logarithms.   Such a cross section, in
which only radiation into a fixed portion of phase
space ($\O$) is specified, was termed ``non-global" by
Dasgupta and Salam, and the associated logarithms
are also called non-global \cite{DS,nonglobal,BanfiMarchSmye}.

In effect, a  non-global definition of energy
flow is not restrictive
enough to limit final states to a specific set of jets, and
non-global logarithms are produced by jets of intermediate energy,
emitted in directions between region $\O$ and
the leading jets.
Thus, interjet energy flow
does not always originate directly from the leading jets, in the absence of
a systematic criterion for suppressing intermediate radiation.
Correspondingly, non-global logarithms
reflect color flow at all scales, and do not
exponentiate in a simple manner.
Our aim in this paper is to formulate a set of observables
for interjet radiation in which non-global logarithms
are replaced by calculable corrections, and which
reflect the flow of color at short distances.
By restricting the sizes of event shapes,
we will limit radiation in
region $\bar{\O}$, while retaining the chosen jet structure.

An important observation  that
we will employ below is that non-global logarithms are not produced
by secondary emissions that are very close to a jet
direction, because a jet of parallel-moving
particles emits soft radiation coherently.  By
fixing the value of an event shape near the
   limit of narrow jets, we avoid
final states with large energies in $\bar{\O}$ away
from the jet axes.
At the same time, we will identify limits in which non-global logarithms
reemerge as leading corrections, and where the
methods introduced to study nongobal effects in Refs.\ 
\cite{DS,nonglobal,BanfiMarchSmye} provide
important insights.

To formalize these observations,
we study below correlated observables for $e^+e^-$
annihilation into two jets.
(In Eq.\ (\ref{event}) $A$ and $B$ denote positron and electron.) In
$e^+e^-$ annihilation dijet
events, the underlying color flow pattern is
simple, which enables us to concentrate  on the
energy flow within the event.  We will
introduce a class of event shapes, $\bar{f}(a)$
      suitable for measuring energy
flow into only part of phase space, with $a$ an
adjustable parameter.
To avoid large non-global
logarithmic corrections we weight events by
   $\exp[-\nu \bar{f}]$, with $\nu$
the Laplace transform conjugate variable.

For the restricted set
of events with narrow jets, energy flow is proportional
to the lowest-order cross section for gluon
radiation into the selected region.  The resummed cross section,
however, remains sensitive to color flow at short distances
through anomalous dimensions associated with coherent
interjet soft emission.  In a sense, our results show that
an appropriate selection of jet events automatically
suppresses nonglobal logarithms, and confirms
the observation of coherence in interjet radiation
\cite{DKTrev,EKSWjet}.

In the next section, we
introduce the event shapes that
we will correlate with energy flow, and describe their relation to
the thrust and jet broadening.
Section 3 contains the details of the
factorization procedure that
characterizes the cross
section in the two-jet limit.  This is followed in Sec.\ 4
by a derivation of the resummation of logarithms of the event shape
and energy flow, following the method introduced
by Collins and Soper \cite{ColSop}.
We then go on in Sec.\ 5 to exhibit
analytic results at leading logarithmic accuracy in $Q_\O/Q$
and next-to-leading logarithm in the event shape.
Section 6 contains
representative numerical results.
We conclude with a summary and a brief
outlook on further applications.

\section{Shape/Flow Correlations}

\subsection{Weights and energy flow in dijet events}

In the notation of Eq.\ (\ref{event}), we will study an event shape
distribution for the process
\be
e^+ + e^- \rightarrow J_1(p_{J_1}) + J_2(p_{J_2})  +
X_{\bar{\O}} \left(\bar{f}\right) + R_\O (Q_\O)\, ,
\label{crossdef}
\ee
at c.m.\ energy $Q\gg Q_\O\gg \Lambda_{\QCD} $.
Two jets with momenta $p_{J_c},\, c = 1,\,2$ emit
soft radiation (only) at wide angles.  Again,
$\O$ is a region between the
jets to be specified below, where the total energy or the transverse
energy $Q_\O$ of the soft radiation is measured,
and $\bar{\O}$ denotes the remaining
phase space (see Fig. \ref{event}).  Radiation into $\bar{\O}$
is constrained by event shape $\bar{f}$.  We
refer to cross sections at fixed values (or transforms) of $\bar{f}$ and
$Q_\O$
as shape/flow  correlations.

To impose the two-jet condition on the states of Eq.\ (\ref{crossdef}) we
choose weights that suppress states with substantial radiation into
$\bar{\O}$
away from the  jet axes.
We now introduce a class of event shapes $\bar{f}$, related
to the thrust, that
enforce the two-jet condition in a natural way.

These event shapes interpolate between and extend the familiar
thrust  \cite{farhi} and jet broadening \cite{broaden1,broaden2},
through an adjustable parameter $a$.
For each state $N$ that defines process (\ref{crossdef}),
we separate $\bar{\O}$ into two regions, $\bar{\O}_c$, $c=1,2$,
containing jet axes, $\hat{n}_c(N)$.  To be specific, we
let $\bar\O_1$ and $\bar\O_2$ be two hemispheres that
cover the entire space except for their intersections with
region $\O$.  Region $\bar\O_1$ is centered on $\hat n_1$,
and $\bar\O_2$ is the opposite hemisphere.
We will specify the
method that determines the jet axes $\hat n_1$ and
$\hat n_2$ momentarily. To identify a meaningful jet, of course, the
total energy
within $\bar\O_1$ should be a large fraction of the available energy,
of the order of $Q/2$
in dijet events.
In $\rm e^+e^-$ annihilation, if there is a well-collimated
jet in $\bar\O_1$ with nearly half the
total energy, there will automatically be one in $\bar{\O}_2$.

We are now ready to define
the contribution from particles in region $\bar\O_c$ to the $a$-dependent
event shape,
\be
\bar f_{\bar{\O}_c}(N,a) =
\frac{1}{\sqrt{s}}\
\sum_{\hat n_i\in \bar \O_c}\ k_{i,\,\perp}^a\, \o_i^{1-a}\,
\left(1-\hat n_i\cdot \hat n_c\right)^{1-a} \, ,
\label{barfdef}
\ee
where $a$ is any real number less than two, and
where $\sqrt{s}=Q$ is the c.m.\ energy.  The
sum is over those particles of state $N$  with direction $\hat n_i$
that flow into
$\bar\O_c$, and their transverse momenta $k_{i,\perp}$ are measured 
relative to $\hat{n}_c$.
The jet axis $\hat n_1$ for jet 1 is identified as that axis
that minimizes the specific thrust-related quantity $\bar 
f_{\bar{\O}_1}(N,a=0)$.
When $\bar{\O}_c$ in Eq.\ (\ref{barfdef}) is extended to all of phase space,
the case $a=0$ is then essentially $1-T$, with $T$ the thrust, while
$a=1$ is related to the
jet  broadening.

Any choice $a<2$ in (\ref{barfdef}) specifies an infrared safe event shape
variable, because the
contribution of any particle $i$ to the event shape behaves as
$\theta_i^{2-a}$ in the collinear limit, $\theta_i=\cos^{-1} (\hat n_i
\cdot \hat
n_c ) \rightarrow 0$.  Negative values of $a$ are clearly permissible, and
the limit $a\rightarrow -\infty$
corresponds to the total cross section.
At the other limit, the factorization and resummation
techniques that we discuss below will apply
only to $a<1$. For
$a> 1$, contributions to the event shape (\ref{barfdef}) from energetic
particles near the jet axis are generically larger than
contributions from soft, wide-angle radiation, or equal for
$a=1$.  When this is the case, the
analysis that we present below must be modified, at
least beyond the level of leading logarithm \cite{broaden2}.

In summary, once $\hat n_1$ is fixed, we have divided the phase space into
three regions:
\begin{itemize}
\item Region $\O$, in which we
measure, for example,
the energy flow,
\item Region $\bar \O_1$, the entire hemisphere centered on
$\hat n_1$, that is, around jet 1, except its intersection with $\O$,
\item Region $\bar \O_2$, the complementary hemisphere, except its
intersection with $\O$.
\end{itemize}
In these terms, we define
the complete event shape variable $\bar f(N,a)$ by
\ba
\bar f(N,a) &=& \bar f_{\bar{\O}_1}(N,a)+\bar f_{\bar{\O}_2}(N,a)\, , 
\label{2jetf}
\ea
with ${\bar f}_{\bar{\O}_c}$, $c=1,2$ given by (\ref{barfdef}) in terms of
the axes $\hat n_1$ of jet 1 and $\hat n_2$ of jet 2.
We will study the correlations of this set of event
shapes with the energy flow into $\O$, denoted as
\be
f(N) =  {1\over \sqrt s}\ \sum_{\hat n_i\in\O} \o_i\, .
\label{eflowdef}
\ee

The  differential cross section
for such dijet events at fixed values of $\bar f$ and $f$ is now
\ba
{d \bar{\sigma}(\varepsilon,\bar{\varepsilon},s, a)\over d \varepsilon
\,d\bar{\varepsilon}\, d\hat n_1}
&=&
{1\over 2s}\ \sum_N\;
|M(N)|^2\, (2\pi)^4\, \delta^4(p_I-p_N) \nonumber\\
&\ & \hspace{10mm} \times
\delta( \varepsilon-f(N))\, \delta(\bar{\varepsilon} -\bar f(N,a))\;
\delta^2  (\hat n_1 -\hat n(N))\, ,
\label{eventdef}
\ea
where we sum over all final states $N$ that contribute to the
weighted event, and where $M(N)$ denotes
the corresponding amplitude for ${\rm e^+e^-}\rightarrow N$.
The total momentum is $p_I$, with $p_I^2=s\equiv Q^2$.
As mentioned in the introduction, for much of our analysis,
we will work with the Laplace
transform of (\ref{eventdef}),
\ba
{d {\sigma}(\varepsilon,\nu,s, a)\over d \varepsilon
\, d\hat n_1}
&=&
\int_0^\infty d{\bar{\varepsilon}}\ {\rm e}^{-\nu{\bar{\varepsilon}}}\
{d \bar{\sigma}(\varepsilon,\bar\varepsilon,s, a)\over d \varepsilon
\,d\bar{\varepsilon}\, d\hat n_1}
\nonumber\\
&=&
{1\over 2s}\ \sum_N\;
|M(N)|^2\, {\rm e}^{-\nu\bar f(N,a)}\
(2\pi)^4\, \delta^4(p_I-p_N) \nonumber\\
&\ & \hspace{10mm} \times
\delta( \varepsilon-f(N))\;
\delta^2  (\hat n_1 -\hat n(N))\, .
\label{transeventdef}
\ea
Singularities of the form
$(1/\bar\varepsilon)\, \ln^n(1/\bar\varepsilon)$ in
the cross section (\ref{eventdef})
give rise to logarithms $\ln^{n+1}\nu$ in the transform
(\ref{transeventdef}).

      Since we are investigating energy
flow in two-jet cross sections, we fix the
constants $\varepsilon$ and $\bar{\varepsilon}$ to be
both much less than unity:
\be
0 < \varepsilon,\bar{\varepsilon} \ll 1.
\label{elasticlim}
\ee
We refer to this as  the elastic limit for the two jets.
In the elastic limit, the dependence of the directions of the
jet axes on soft radiation is weak.  We will return to
this dependence below.
Independent of soft radiation, we can
always choose our coordinate system such
that the
transverse momentum of jet 1 is
zero,
\be
p_{J_1,\, \perp} = 0\, ,
\ee
with $\vec p_{J_1}$ in the $x_3$ direction.  In the limit $\bar
\varepsilon, \varepsilon\rightarrow 0$, and in the overall c.m.,
$p_{J_1}$ and $p_{J_2}$ then approach light-like vectors in the plus and
minus directions:
\ba
p_{J_1}^\mu &\rightarrow&  \left(\sqrt{\frac{s}{2} },0^-,0_\perp \right)
\nonumber\\
p_{J_2}^\mu &\rightarrow&  \left(0^+,\sqrt{\frac{s}{2} },0_\perp \right)\, .
\label{lightlike}
\ea
As usual, it is convenient to work in light-cone coordinates,
$p^\mu  =  \left( p^+ , p^-, p_\perp \right)$, which we normalize as
$p^\pm  =  (1/\sqrt{2})(p^0 \pm p^3)$.
For small $\varepsilon$ and $\bar{\varepsilon}$, the cross section
(\ref{eventdef}) has
corrections in $\ln (1/\varepsilon)$ and
$\ln (1/\bar{\varepsilon})$, which we will organize in the following.

\subsection{Weight functions and jet shapes}

In Eq.\ (\ref{barfdef}), $a$ is a parameter that allows us to study
various event
shapes within the same formalism; it helps to control the
approach to the two-jet limit.   As noted above,
$a< 2$ for infrared safety, although the factorization
that we will discuss below applies beyond leading logarithm
only to $1>a>-\infty$.  A
similar weight function with a non-integer power has been discussed in
a related context for $2>a>1$ in
\cite{manoharwise}.
To see how the parameter $a$ affects the shape of the jets, let us
reexpress
the weight function for jet 1 as
\ba
\bar f_{\bar{\O}_1}(N,a) = \frac{1}{\sqrt{s}}\
\sum_{\hat n_i\in \bar \O_1} \o_i \sin^a \theta_i \left( 1 -
\cos \theta_i  \right)^{1-a}, \label{fbarexp}
\ea
where $\theta_i$ is the angle of the momentum of final state
particle $i$ with respect to jet axis $\hat n_1$.
As $a \rightarrow 2$ the weight vanishes only  very slowly for
$\theta_i\rightarrow 0$, and at fixed $\bar f_{\bar{\O}_1}$, the
jet becomes very narrow. On the other hand, as $a \rightarrow -
\infty$, the event
shape vanishes more and more rapidly in the forward direction, and the
cross
section at fixed $\bar f_{\bar{\O}_1}$ becomes more
and more inclusive in the radiation into $\bar{\O}_1$.

In this paper, as in Ref.\ \cite{BKS1},
we seek to control corrections in the single-logarithmic variable
$\alpha_s(Q) \ln (1/\varepsilon)$,
with $\varepsilon=Q_\O/Q$.  Such a resummation is most
relevant when
\be
\alpha_s(Q) \ln \left({1 \over \varepsilon}\right) \ge 1 \rightarrow
\varepsilon \le  \exp\left({- 1\over \alpha_s(Q) }\right)\, .
\label{einequal}
\ee
Let us compare these logarithms to non-global
effects in shape/flow correlations.
At $\nu=0$ and for $a\rightarrow -\infty$,
the cross section becomes inclusive outside $\O$.  As we show below,
the non-global logarithms discussed in Refs.\ \cite{BKS1, DS}
appear in shape/flow correlations as logarithms of the form
$\alpha_s(Q)\, \ln(1/(\varepsilon \nu))$, with $\nu$ the moment variable
conjugate to the event shape.  To treat these logarithms
as subleading for small $\varepsilon$ and (relatively)
large $\nu$, we require that
\ba
\alpha_s(Q)\, \ln \left({1 \over \varepsilon \nu}\right) < 1  \rightarrow
\varepsilon >
{1\over\nu}\;
\exp\left({- 1\over \alpha_s(Q) }\right)     \, .
\label{enuinequal}
\ea
For large $\nu$, there is a substantial range of $\varepsilon$
in which both (\ref{einequal}) and (\ref{enuinequal}) can
hold.  When $\nu$ is large, moments of the
correlation are dominated precisely by events with
strongly two-jet energy flows, which is the natural
set of events in which to study the influence of color
flow on interjet radiation.  (The peak of
the thrust cross section is at $(1-T)$ of order one-tenth
at LEP energies, corresponding to $\nu$ of order ten,
so the requirement of large $\nu$ is not overly restrictive.)
  In the next subsection, we
show how the logarithms of $(\varepsilon \nu )^{-1}$ emerge in a
low order example. This analysis
also assumes that $a$ is not large in absolute value. The event shape
at fixed angle decreases exponentially with $a$, and we
shall see that higher-order corrections can be proportional to $a$.
We always treat $\ln \nu$ as much larger than $|a|$.

\subsection{Low order example}
\label{sec:loe}

\begin{figure} \center
\includegraphics*{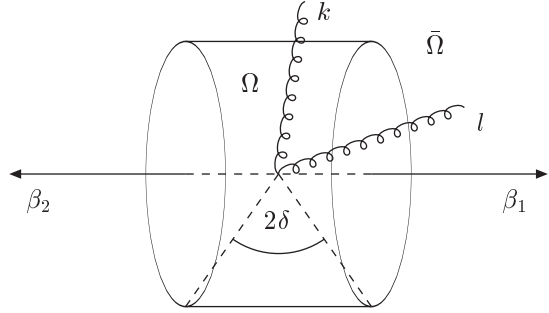}
\caption{\label{kinematics} A kinematic configuration that gives rise to
the
non-global logarithms. A soft gluon with momentum $k$ is radiated
into the region $\Omega$, and an
energetic gluon with momentum $l$ is radiated into $\bar\Omega$.
Four-vectors
$\beta_1$ and $\beta_2$, define the directions of jet 1 and jet 2,
respectively.}
\end{figure}

In this section, we check the general ideas developed above with
the concrete
example of a two-loop cross section for the process
(\ref{crossdef}). This is the lowest order in which a
non-global logarithm occurs, as observed in \cite{DS}. We
normalize this cross section to the Born cross section for
inclusive dijet production. A similar
analysis for the same geometry has been carried out in \cite{DS} and
\cite{nonglobalAS}.

The kinematic configuration we consider is shown in Fig. \ref{kinematics}.
Two fast partons, of velocities
    $\vec{\beta_1}$ and $\vec{\beta_2}$, are treated in eikonal
approximation.
In addition, gluons are emitted into the final state.
A soft gluon with momentum $k$ is radiated into region $\Omega$ and
an energetic gluon with momentum
$l$ is emitted into the region $\bar{\Omega}$.
We consider the cross section at fixed energy,
$\o_k\equiv \varepsilon\sqrt{s}$.
As indicated above, non-global logarithms arise from
strong ordering of the energies of the gluons,
which we choose as $ \ol \gg \ok $.
In this region, the gluon $l$ plays the role of a ``primary"
emission, while $k$ is a ``secondary" emission.

\begin{figure} \center
\includegraphics*{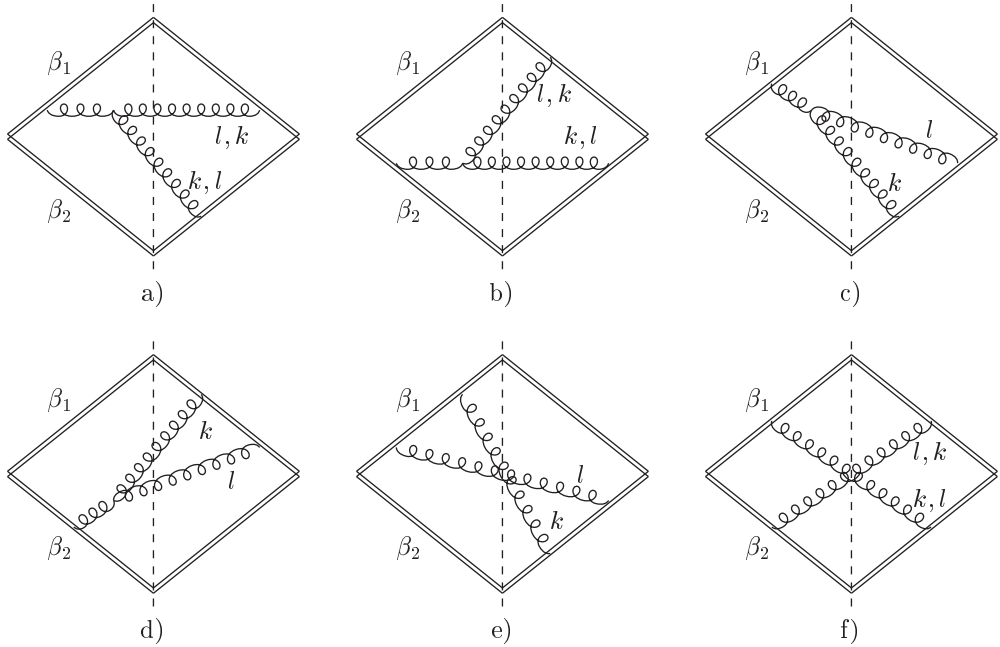}
\caption{\label{diagrams} The relevant two-loop cut diagrams
corresponding to the emission of
two real gluons in the final state contributing to the eikonal cross
section.
The dashed line represents the final state, with
contributions to the amplitude
to the left, and to the complex conjugate amplitude to the right.}
\end{figure}

For our calculation, we take the angular region $\Omega$
to be a ``slice" or ``ring'' in polar
angle of width $2 \delta$, or
equivalently, (pseudo) rapidity interval $(-\eta,\eta)$, with
\be \label{rapidity}
\Delta \eta =2\eta= \ln\left(\frac{1+\sd}{1-\sd}\right)\, ,
\label{deltaeta1}
\ee
The lowest-order diagrams for
this process are those shown
in Fig. \ref{diagrams}, including distinguishable diagrams
in which the momenta $k$ and $l$ are interchanged.

The diagrams of Fig.\ \ref{diagrams}
give rise to color structures $C_F^2$ and $C_FC_A$,
but terms proportional to $C_F^2$  may be associated with a factorized
contribution to the cross section, in which the
gluon $k$ is emitted coherently by the combinations
of the gluon $l$ and the eikonals.
To generate the $C_FC_A$ part, on the other hand, gluon $k$
must ``resolve"
gluon $l$ from the eikonal lines,
giving a result that depends on
the angles between $\vec l$ and the
eikonal directions.

The computation of the diagrams is outlined
in Appendix \ref{eikapp}; here we quote the results.  We adopt the notation
$c_l\equiv \cos\theta_l,\,s_l \equiv \sin\theta_l$, with $\theta_l$
the angle of momentum $\vec l$ measured relative
to $\vec \beta_1$, and similarly for
$k$.   We take, as indicated above, a Laplace transform
with respect to the shape variable, and
identify the logarithm in the conjugate variable $\nu$.
We find that the logarithmic $C_FC_A$-dependence
of Fig.\ \ref{diagrams} may be written as a dimensionless
eikonal cross section in
terms of one energy and two polar angular integrals as
\bea
\label{ps0}
{d\sigma_{\rm eik}\over d\, \varepsilon}
    & = & C_F C_A \left(\frac{\alpha_s}{\pi}\right)^2 \,
\frac{1}{\varepsilon} \,
\int_{-\sd}^{\sd}
\mathrm{d} c_k \, \int_{\sd}^{1}
\mathrm{d} c_l \, \int_{\varepsilon \sqrt{s}}^{\sqrt{s}} \frac{\mathrm{d}
\ol}{\ol} \, e^{-\nu \, \ol \,
(1-c_l)^{1-a} \, s_l^a / Q}
\nonumber \\
& & \times\ \left[ \frac{1}{c_k + c_l} \, \frac{1}{1+c_k}
\left(\frac{1}{1+c_l} + \frac{1}{1-c_k}\right) - \frac{1}{s_k^2} \,
\frac{1}{1+c_l} \right]\, .
\eea
In this form, the absence of collinear singularities
in the $C_FC_A$ term at $\cos\theta_l=+ 1$ is manifest, independent of $\nu$.
Collinear singularities in the $l$ integral completely
factorize from the $k$ integral, and are proportional
to $C_F^2$.
The logarithmic dependence
on $\varepsilon$ for $\nu > 1$ is readily found to be
\ba
\label{psa}
{d\sigma_{\rm eik}\over d\, \varepsilon} = C_F C_A
\left(\frac{\alpha_s}{\pi}\right)^2 \, \frac{1}{\varepsilon} \,
\ln\left(\frac{1}{\varepsilon \nu}\right)\, C(\Delta \eta)\, ,
\ea
where  $C(\Delta \eta)$ is a finite function of the
angle $\delta$, given explicitly in
Appendix \ref{eikapp}.

We can contrast  this result to what happens when $\nu=0$,
that is, for an inclusive,  non-global cross section.  In this case,
recalling that $\varepsilon=Q_\O/Q$,
we find in place of Eq.\ (\ref{psa}) the non-global logarithm
\ba
\label{psb}
{d\sigma_{\rm eik}\over d\, \varepsilon} = C_F C_A
\left(\frac{\alpha_s}{\pi}\right)^2 \, \frac{1}{\varepsilon} \,
\ln\left(\frac{Q}{Q_\O}\right)\, C(\Delta \eta)\, .
\ea
As anticipated, the effect of the transform is to
replace the non-global logarithm in $Q/Q_\O$, by a logarithm
of $1/( \varepsilon \nu)$.  We are now ready to generalize this
result, starting from the factorization properties of
the cross section near the two-jet limit.

\section{Factorization of the Cross Section}

In this section we study the factorization of the
correlations (\ref{eventdef}).  The analysis
is based on a general approach
that begins with the all-orders treatment of singularities
in perturbative cross sections \cite{power,GStasi},
and derives factorization from the analyticity and
gauge properties of high energy Green functions and cross
sections \cite{pQCD}. The functions
that appear in factorized cross sections are expressible
in terms of QCD matrix elements \cite{pdfdef}, and
the matrix elements that we will encounter are familiar
from related analyses for heavy quark and jet production
\cite{hqjet}.  We refer in several places below
to standard arguments discussed in more detail
in \cite{GStasi,pQCD}.
The aim
of this section, and the reason why a
careful analysis is necessary, is to identify
the specific dimensionless
combinations of kinematic variables
on which the factorized matrix elements may depend.
We will use these dependences  in the following section,
when we discuss the resummation properties of
our correlations.

\subsection{Leading regions near the two-jet limit}

In order to resum logarithms of $\varepsilon$ and $\bar{\varepsilon}$
(or equivalently $\nu$, the Laplace conjugate of $\bar{\varepsilon}$) we
have first to
identify their origin in momentum space when
$\varepsilon,\bar{\varepsilon}\rightarrow 0$. Following the procedure
and terminology
of \cite{power}, we identify ``leading regions" in
the momentum integrals of cut diagrams, which can give rise
to logarithmic enhancements of the
cross section associated with lines approaching the
mass shell.  Within these regions, the lines of a cut diagram
fall into the following subdiagrams:

\begin{itemize}
\item A hard-scattering, or ``short-distance" subdiagram $H$, where all
components of line momenta are far off-shell, by order $Q$.
\item Jet subdiagrams, $J_1$ and $J_2$, where energies are
fixed and momenta are collinear
to the outgoing primary partons and the jet
directions that emerge from the hard scattering. (For
$\varepsilon=\bar{\varepsilon}=0$,
the sum of all energies in each jet is one-half the total energy.)
To characterize the momenta of the lines within the jets,
we introduce a scaling variable, $\lambda\ll 1$.  Within
jet 1, momenta $\ell$ scale as $(\ell^+ \sim Q,\ell^-\sim \lambda
Q,\ell_\perp \sim \lambda^{1/2}Q)$.
\item A soft subdiagram, $S$ connecting the jet functions $J_1$ and
$J_2$, in which the components of
   momenta $k$ are  small compared to $Q$ in all components,
scaling as $(k^\pm \sim \lambda Q,k_\perp \sim \lambda Q)$.
\end{itemize}

An arbitrary final state $N$ is the
union of substates associated with these subdiagrams:
\be
N=N_s \oplus N_{J_1} \oplus N_{J_2}\, .
\ee
As a result, the  event shape $\bar f$ can
also be written as a sum of contributions from the soft
and jet subdiagrams:
\ba
\bar f(N,a) &=& \bar f^N(N_s,a) + \bar f^N_{\bar{\O}_1}(N_{J_1},a) +
\bar f^N_{\bar{\O}_1}(N_{J_2},a)\, .
\label{fbarf}
\ea
The superscript $N$ reminds us that the contributions of
final-state particles associated with
the soft and jet functions depend implicitly on the
full final state, through the determination of the
jet axes, as discussed in Sec.\ 2.
In contrast, the energy flow weight, $f(N)$, depends only on
particles emitted at wide angles, and is hence insensitive
to collinear radiation:
\ba
f(N) = f(N_s)\, .
\ea

When we sum over all diagrams
that have a fixed final state, the contributions from these leading regions
may be factorized into a set of functions, each of which corresponds to
one of the generic hard, soft and jet subdiagrams.  The arguments for this
factorization at
leading power have been discussed extensively \cite{ColSop,pQCD,ColSte}.
The cross section becomes a convolution in
$\bar \varepsilon$, with the sums over states linked
by the delta function which fixes $\hat n_1$, and by momentum
conservation,
\ba
{d \bar{\sigma}(\varepsilon,\bar{\varepsilon},s,a)\over d \varepsilon
\,d\bar{\varepsilon}\, d\hat n_1}
&=& {d \sigma_0 \over d\hat{n}_1}\
H(s,\hat{n}_1)\
\sum_{N_s,N_{J_c}}\;
\int d\bar{\varepsilon}_s\, {\cal S}(N_s)\, \delta(\varepsilon-f(N_s))\,
\delta(\bar{\varepsilon}_s-\bar{f}^N(N_s,a))\nonumber\\
&\ & \quad \times \prod_{c=1}^2\,
   \int  d\bar{\varepsilon}_{J_c}\, {\cal J}_c (N_{J_c}) \,
\delta(\bar{\varepsilon}_{J_c}-\bar{f}^N_{\bar{\O}_c}(N_{J_c},a))\nonumber\\
&\ & \hspace{15mm} \times\
(2\pi)^4\, \delta^4(p_I-p(N_{J_2})-p(N_{J_1})-p(N_s))
\nonumber\\
&\ & \hspace{15mm} \times\
\delta^2(\hat n_1 -\hat n(N))\;
\delta(\bar{\varepsilon}-\bar{\varepsilon}_{J_1}-\bar{\varepsilon}_{J_2}-\bar{
\varepsilon}_s)
\nonumber\\
&=& {d \sigma_0 \over d\hat{n}_1}\; \delta(\varepsilon)\,
\delta(\bar{\varepsilon})+{\cal O}(\alpha_s)\, .
\label{sigmafact}
\ea
Here  $d\sigma_0/d\hat{n}_1$ is the Born
cross section for the production of a single
particle (quark or antiquark) in direction
$\hat{n}_1$, while the short-distance function
$H(s,\hat{n}_1)=1+{\cal O}(\alpha_s)$, which
    describes corrections to the hard scattering,
is an expansion in $\alpha_s$ with finite coefficients.
The functions ${\cal J}_c (N_{J_c}),\ {\cal S}(N_s)$
describe
the internal dynamics of the jets and wide-angle soft
radiation, respectively.  We will specify these functions below.
We have suppressed their dependence on a factorization scale.
Radiation at wide angles from the jets will be well-described
by our soft functions ${\cal S}(N_s)$, while
we will construct
the jet functions ${\cal J}_c (N_{J_c})$ to be independent of
$\varepsilon$, as  in
Eq.\ (\ref{sigmafact}).

So far, we have specified our sums over states in Eq.\ (\ref{sigmafact})
only when all lines in $N_s$ are
soft, and all lines in $N_{J_c}$ have momenta that are collinear, or
nearly collinear
to $p_{J_c}$.   As $\varepsilon$ and $\bar{\varepsilon}$ vanish, these are the
only final-state momenta that are kinematically possible.
Were we to restrict ourselves to these configurations
only, however, it would not be straightforward to make the individual
sums over $N_s$ and $N_{J_c}$ infrared safe.  Thus, it is necessary to
include soft partons in $N_s$ that are emitted near the jet directions,
and soft partons in the $N_{J_c}$ at wide angles.
We will show below how to define the functions ${\cal J}_c (N_{J_c}),\ {\cal
S}(N_s)$
so that they generate factoring, infrared safe functions that
avoid double counting.
We know on the basis of the arguments of Refs.\ \cite{ColSop,pQCD,ColSte}
that corrections to
the factorization of soft from jet functions are suppressed by
powers of the weight functions $\varepsilon$ and/or $\bar \varepsilon$.

\subsection{The factorization in convolution form}

Although formally factorized, the jet and soft functions
in Eq.\ (\ref{sigmafact}) are still  linked in a potentially complicated
way through their dependence on the jet
axes.  Our strategy is to simplify this complex dependence
to a simple convolution in contributions to $\bar\varepsilon$,
accurate to leading power in $\varepsilon$ and $\bar\varepsilon$.

First, we note that the cross section of Eq.\ (\ref{sigmafact})
is singular for vanishing $\varepsilon$ and $\bar{\varepsilon}$, but is a
smooth function of $s$ and $\hat{n}_1$.  We may therefore make any
approximation that changes $s$ and/or $\hat{n}_1$ by an amount
that vanishes as a power of $\varepsilon$ and $\bar{\varepsilon}$ in the
leading regions.

Correspondingly, the amplitudes for jet $c$ are singular in
$\bar{\varepsilon}_{J_c}$,
but depend smoothly on the jet energy and direction, while
the soft function is singular in both $\varepsilon$ and
$\bar{\varepsilon}_s$,
but depends smoothly on the jet directions.  As a result,
at fixed values of $\varepsilon$ and $\bar\varepsilon$ we
may approximate the jet directions and energies by
their values at $\varepsilon=\bar{\varepsilon}=0$ in the soft and jet
functions.

Finally, we may make any approximation that affects
the value of $\varepsilon$ and/or $\bar\varepsilon_{J_c}$ by
amounts that vanish faster than linearly for $\bar\varepsilon\rightarrow 0$.
It is at this stage that we will require that $a<1$.

With these observations in mind,
we enumerate the replacements
and approximations by which we
reduce Eq.\ (\ref{sigmafact}), while retaining leading-power accuracy.

\begin{enumerate}

\item  To simplify the definitions  of the jets in Eq.\ (\ref{sigmafact}),
we make the replacements $\bar{f}^N_{\bar{\O}_c}(N_{J_c},a)
\rightarrow \bar f_c(N_{J_c},a)$ with
\ba
\bar f_c(N_{J_c},a) \equiv
\frac{1}{\sqrt{s}} \sum_{{\rm all}\
\hat n_i
\in N_{J_c} }\
k_{i,\,\perp}^a\, \o_i^{1-a}\, \left(1-\hat n_i\cdot \hat n_c
\right)^{1-a}\, .
\label{fbar2jet1}
\ea
The jet weight function $\bar{f}_c(N_{J_c},a)$ now depends only on particles
associated with $N_{J_c}$.
   The contribution to $\bar{f}_c(N_{J_c},a)$
from particles within region $\bar{\O}_c$,
is exactly the same here as in the weight (\ref{barfdef}),
but we now include particles
in all other directions.
     In this way, the independent sums over final states of the
jet amplitudes will be naturally infrared
safe.  The value of $\bar f_c(N_{J_c},a)$ differs from
the value of $\bar{f}^N_{\bar{\O}_c}(N_{J_c},a)$, however, due to
radiation outside
$\bar\O_c$, as indicated by the new subscript.  This radiation is hence at
wide angles to the jet axis.  In the elastic limit (\ref{elasticlim}), it is
also constrained to be soft.  Double counting in contributions
to the total event shape, $\bar f(N,a)$, will be avoided by an appropriate
definition
of the soft function below.
The sums over states are still not yet fully independent,
however, because the jet directions $\hat n_c$ still depend
on the full final state $N$.

\item
Next, we turn our attention to the condition that fixes the jet
direction $\hat n_1$.
Up to corrections in the orientation of $\hat n_1$
that vanish as powers of $\varepsilon$ and $\bar{\varepsilon}$,
we may neglect the dependence of $\hat{n}_1$
on $N_s$ and $N_{J_2}$:
\be
\delta(\hat n_1-\hat n(N))  \rightarrow  \delta(\hat n_1 -\hat n(N_{J_1}))\, .
\label{nnJone}
\ee
In Appendix \ref{approxapp}, we show that
this replacement also leaves the value of
$\bar\varepsilon$ unchanged, up to corrections that vanish as
$\bar\varepsilon^{2-a}$.  Thus, for $a<1$, (\ref{nnJone}) is
acceptable to leading power.
For $a<1$, we can therefore
identify the direction of jet 1 with $\hat{n}_1$.
These approximations simplify Eq.\ (\ref{sigmafact})
by eliminating the implicit dependence of
the jet and soft weights on the full final state.  We may
now treat $\hat n_1$ as an independent vector.

\item  In the leading regions, particles
that make up each final-state jet are associated with states $N_{J_c}$,
while $N_s$  consists of soft particles only.
In the momentum conservation delta function, we
can neglect the four-momenta of lines in $N_s$,
whose energies all vanish as $\varepsilon,\bar{\varepsilon}\rightarrow 0$:
\be
\delta^4(p_I-p(N_{J_2})-p(N_{J_1})-p(N_s))
\rightarrow
\delta^4(p_I-p_{J_2}-p_{J_1}).
\ee

\item
Because the cross section is a smooth function of
the jet energies and directions, we may also
neglect the masses of the jets within the
momentum conservation delta function, as in
Eq.\ (\ref{lightlike}).  In this approximation,
we derive in the c.m.,
\ba
\delta^4(p_I-p_{J_2}-p_{J_1})
&\rightarrow&
\delta(\sqrt{s} - \o(N_{J_1})-\o(N_{J_2}))
\, \delta(|\vec p_{J_1}|-|\vec p_{J_2}|)
\, {1\over |\vec p_{J_1}|^2}
\, \delta^2(\hat n_1 + \hat n_2)\nonumber\\
&\rightarrow& {2\over s}\,
\delta\left({\sqrt{s}\over 2} - \o(N_{J_1})\right)
\, \delta\left({\sqrt{s}\over 2} - \o(N_{J_2})\right)
\, \delta^2(\hat n_1 + \hat n_2)\, .
\ea
Our jets are now back-to-back:
\be
\hat n_2 \rightarrow -\hat n_1\, .
\label{fbar2soft}
\ee

\end{enumerate}

Implementing these replacements and approximations for $a<1$,
we rewrite the cross section Eq.\
(\ref{sigmafact}) as
\ba
{d \bar{\sigma}(\varepsilon,\bar{\varepsilon},s,a)\over d \varepsilon\,
d\bar{\varepsilon}\, d\hat n_1}
&=&
{d \sigma_0 \over d\hat{n}_1}\
H(s,\hat{n}_1,\mu)\;
\int  d\bar{\varepsilon}_s\,
\bar{S}(\varepsilon,\bar{\varepsilon}_s,a,\mu) \,
\nonumber\\
&\ & \times
\prod_{c=1}^2\, \int  d\bar{\varepsilon}_{J_c}\,
\bar{J}_c(\bar{\varepsilon}_{J_c},a,\mu)\,
\delta(\bar{\varepsilon}- \bar{\varepsilon}_{J_1}-\bar{\varepsilon}_{J_2}-
\bar{\varepsilon}_s)\, ,
\label{factor}
\ea
with (as above) $H=1+{\cal O}(\alpha_s)$.  Referring to
the notation of Eqs.\ (\ref{sigmafact}) and (\ref{fbar2jet1}),
the functions $\bar{S}$ and $\bar{J}_c$ are:
\ba
\bar{S}(\varepsilon,\bar{\varepsilon}_s,a,\mu)
&=&
\sum_{N_s}\; {\cal S}(N_s,\mu)\, \delta(\varepsilon-f(N_s)) \,
\delta(\bar{\varepsilon}_s-\bar f(N_s,a))
\label{firstSdef}
\\
\bar{J}_c(\bar{\varepsilon}_{J_c},a,\mu)
&=&
\frac{2}{s}  (2\pi)^6\,   \sum_{N_{J_c}}
{\cal J}_c(N_{J_c},\mu) \, \delta(\bar{\varepsilon}_{J_c}-\bar 
f_c(N_{J_c},a))\,
\delta\left({\sqrt{s}\over 2} - \o(N_{J_c})\right)\,
\delta^2(\hat n_1 \pm \hat n(N_{J_c})),
\nonumber\\
\label{firstJdef}
\ea
with the plus sign in the angular delta function
for jet 2, and the minus for jet 1. The weight functions for the jets are
given by Eq.\
(\ref{fbar2jet1}) and induce dependence on
the parameter $a$.   We have introduced the factorization scale
$\mu$, which we set equal to the
renormalization scale.

We note that we must construct the soft functions $\bar{S}(N_s,\mu)$
to cancel the contributions of final-state particles from
each of the $\bar{J}_c(N_{J_c},\mu)$ to the weight $\varepsilon$,
as well as the contributions of the jet functions to $\bar\varepsilon$
from soft radiation outside their respective regions $\bar\O_c$.
Similarly, the jet amplitudes
must be constructed to include collinear enhancements only in their
respective jet directions.  Explicit constructions that satisfy these
requirements will be
specified in the following subsections.

To disentangle the convolution in (\ref{factor}), we take Laplace
moments with respect to $\bar{\varepsilon}$:
\ba
     {d \sigma(\varepsilon,\nu,s,a)\over d \varepsilon \,d\hat n_1}
& = &  \int_0^\infty d\bar{\varepsilon}\, e^{- \nu \,\bar{\varepsilon}}\,
{d \bar{\sigma}(\varepsilon,\bar{\varepsilon},a)\over d \varepsilon\,
d\bar{\varepsilon}\,
d\hat n_1}
\label{trafo} \nonumber
\\
& = & {d \sigma_0 \over d\hat{n}_1}\
H(s,\hat{n}_1,\mu)\;  S(\varepsilon,\nu,a,\mu) \,\prod_{c=1}^2\,
J_c(\nu,a,\mu).
\label{trafosig}
\ea
Here and below  unbarred quantities are the transforms in
$\bar{\varepsilon}$,
and barred quantities denote untransformed functions.
\be
S(\varepsilon,\nu,a,\mu) =   \int_0^\infty d\bar{\varepsilon}_s \,e^{- \nu
\,\bar{\varepsilon}_s}
\bar{S}(\varepsilon,\bar{\varepsilon}_s,a,\mu),
\label{trafodef}
\ee
and similarly for the jet functions.

\begin{figure}[htb]
\begin{center}
\epsfig{file=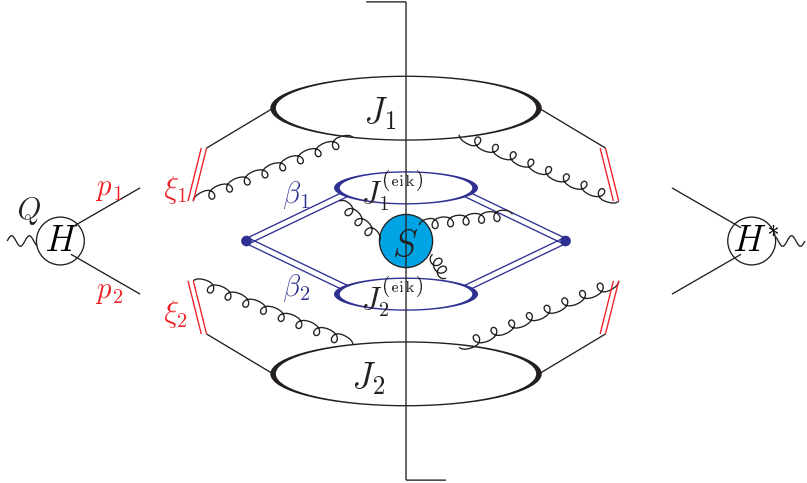,height=7cm,clip=0}
\caption{Factorized cross section (\ref{factor})
after the application of Ward identities. The vertical line
denotes the final state cut.} \label{factorized}
\end{center}
\end{figure}

In the following subsections, we give explicit constructions for the functions
participating in the factorization formula (\ref{factor}),
which satisfy the requirement of infrared safety,
and avoid double counting.
An  illustration of the cross section factorized into these functions is
shown in Fig.\ \ref{factorized}.
    As discussed above,
non-global logarithms will emerge when $ \varepsilon \nu$ becomes small
enough.

\subsection{The short-distance function}
\label{sec:sdf}

The power counting described in \cite{power} shows
that in  Feynman  gauge
the subdiagrams of Fig.\ \ref{factorized} that contribute to
$H$ in Eq.\ (\ref{factor}) at leading
power in $\varepsilon$ and
$\bar \varepsilon$ are connected to
each of the two jet subdiagrams
   by a single on-shell
quark  line,  along with a possible set of on-shell, collinear gluon lines
that carry scalar polarizations.
The hard subdiagram is
not connected directly to the soft subdiagram in any leading region.

The couplings of the scalar-polarized gluons that connect the jets with
short-distance subdiagrams
may be simplified with the help of Ward
identities (see, e.\,g.\ \cite{pQCD}).  At each order of
perturbation theory, the coupling of scalar-polarized gluons
from either jet to the short-distance function is equivalent
to their coupling to a path-ordered exponential of
the gauge field, oriented in any direction that is not
collinear to the jet.  Corrections are infrared safe, and
can be absorbed into the short-distance function.
Let $h(p_{J_c},\hat{n}_1,{\mathcal{A}})$ represent
the set of all short-distance contributions to diagrams
that couple any number of scalar-polarized gluons to the jets,
in the amplitude for the production of any final state.  The argument
   ${\mathcal{A}}$ stands for the fields that create the
scalar-polarized gluons linking the short-distance function
to the jets.
On a diagram-by-diagram basis, $h$ depends on
the momentum of each of the scalar-polarized gluons.
After the sum over all diagrams, however,
   we can make the replacement:
\be
h(p_{J_c},\hat{n}_1,{\mathcal{A}}^{({\rm q,\bar{q}})})
\rightarrow
\Phi^{({\rm \bar{q}})}_{\xi_2} (0,-\infty;0)\, h_2(p_{J_c},\hat{n}_1,\xi_c)\,
\Phi^{({\rm q})}_{\xi_1} (0,-\infty;0) \, ,
\ee
where $h_2$ is a short-distance function that depends only on the
total momenta $p_{J_1}$ and $p_{J_2}$.  It also depends on vectors $\xi_c$
that characterize  the path-ordered
exponentials $\Phi(0,-\infty;0)$:
\be
\Phi^{(\rm f)}_{\xi_c} (0,-\infty;0)  =  P e^{-i g \int_{-\infty}^{0} d
\lambda\; \xi_c \cdot {\mathcal{A}}^{(\rm f)} (\lambda \xi_c )}\, ,
\label{patho}
\ee
where the superscript $(\rm f)$ indicates that the vector potential
takes values in
representation $\rm f$, in our case the representation of a quark or
antiquark.
These operators will be associated with
gauge-invariant definitions of the jet functions below.
To avoid spurious collinear singularities,
we choose the vectors $\xi_c$, $c=1,2$, off the light cone.
   In the full cross section (\ref{trafosig}) the
$\xi_c$-dependence cancels, of course.

The dimensionless short-distance function $H=\left|h_2\right|^2$ in
Eq.\ (\ref{factor})
depends on $\sqrt{s}$ and $p_{J_c}\cdot \xi_c$, but not
on any variable that vanishes with $\varepsilon$ and $\bar{\varepsilon}$:
\be
H(p_{J_c},\xi_c,\hat{n}_1,\mu) =  H \left(
\frac{\sqrt{s}}{\mu},\frac{p_{J_c} \cdot
\xic}{\mu},\hat{n}_1,\as(\mu) \right)\, ,
\label{harddef}
\ee
where
\be
\xic \equiv \xi_c / \sqrt{|\xi_c^2|}\, .
\ee
Here we have observed that each diagram is independent of the overall
scale of the eikonal vector $\xi^\mu_c$.

\subsection{The jet functions}\label{jets}

The jet functions and the soft functions in Eq.\ (\ref{factor})
can be defined in terms of specific matrix elements, which
absorb the relevant contributions to leading regions in
the cross section, and which are infrared safe.
Their perturbative expansions
   specify the functions ${\cal S}$ and ${\cal J}_c$ of
Eq.\ (\ref{firstJdef}).  We begin with
our definition of the jet functions.

The jet functions, which absorb enhancements collinear to the two
outgoing particles produced in the primary hard scattering, can be
defined in terms of matrix elements
in a manner reminiscent of parton distribution or decay functions 
\cite{pdfdef}.
To be specific, we consider the
quark jet function:
\ba
\bar{J'}_c^\mu (\bar{\varepsilon}_{J_c},a,\mu)
&=&
   {2\over s}\, \frac{(2\pi)^6}{\Ncol} \; \sum\limits_{N_{J_c}}
{\rm Tr} \left[\gamma^\mu
\left<0 \left|
\Phi^{(\rm q)}_{\xi_c}{}^\dagger(0,-\infty;0) q(0) \right| N_{J_c} \right>
   \left< N_{J_c}\left| \bar{q}(0) \Phi^{(\rm q)}_{\xi_c}(0,-\infty;0)
\right| 0 \right>\right] \nonumber \\
& &\, \hspace{15mm} \times \, \delta(\bar{\varepsilon}_{J_c}-\bar
f_c(N_{J_c},a))\,
\delta\left({\sqrt{s}\over 2} - \o(N_{J_c})\right)\,
\delta^2(\hat n_c - \hat n(N_{J_c}))
\, ,
\label{jetdef}
\ea
where $\Ncol$ is the number of colors, and
where $\hat n_c$ denotes the direction of the momentum of
   jet $c$, Eq.\ (\ref{firstJdef}),
with $\hat{n}_2 = - \hat n_1$.
    $q$ is the quark field, $\Phi_{\xi_c}^{(\rm q)}(0,-\infty;0)$
a path-ordered exponential in the notation of (\ref{patho}),
and the trace is taken over color and Dirac indices.
We have chosen the normalization so that the
jet functions $\bar{J}'{}^\mu$ in (\ref{jetdef}) are dimensionless
and begin at lowest order with
\ba
\bar{J'}_c^\mu{}^{(0)} (\bar{\varepsilon}_{J_c},a,\mu) = \beta_{c}^\mu\,
\delta({\bar \varepsilon}_{J_c})\, ,
\label{norm}
\ea
with $\beta_c^\mu$
the lightlike velocities corresponding to the jet momenta in Eq.\ 
(\ref{lightlike}):
\be
\beta_1^\mu=\delta_{\mu +}\ , \quad \beta_2^\mu = \delta_{\mu -}\, .
\label{betadef}
\ee
The scalar jet functions of Eq.\
(\ref{firstJdef}) are now obtained by projecting out
the component of $J'_c{}^\mu$ in the jet direction:
\be
\bar{J}_c (\bar{\varepsilon}_{J_c},a,\mu) =  \bar{\beta}_c \cdot
\bar{J'}_c
(\bar{\varepsilon}_{J_c},a,\mu) = \delta(\bar{\varepsilon}_{J_c}) +{\cal
O}(\alpha_s)\, ,
\label{Jnorm}
\ee
where
$\bar{\beta}_1=\beta_2$, $\bar{\beta_2}=\beta_1$ are the lightlike vectors
in the directions opposite to $\beta_1$ and $\beta_2$, respectively.
By construction, the
$\bar{J}_c$ are linear in $\bar{\beta}_c$.

To resum the jet functions in the variables $\bar{\varepsilon}_{J_c}$,
it is convenient to reexpress the weight functions
(\ref{fbar2jet1}) in combinations of light-cone momentum
components that are invariant under boosts in the $x_3$ direction,
\ba
\bar{f}_1\left(N_{J_1},a\right) & = & \frac{1}{s^{1-a/2}}
\sum_{\hat n_i \in N_{J_1} }\
k_{i,\,\perp}^a\, \left(2 p_{J_1}^+k_i^-\right)^{1-a},
   \label{fbarLC1}
\nonumber \\
\bar{f}_2\left(N_{J_2},a \right) & = & \frac{1} {s^{1-a/2}}
\sum_{\hat n_i \in N_{J_2} }\
k_{i,\,\perp}^a\, \left(2 p_{J_2}^-k_i^+\right)^{1-a}.
\label{fbarLC2}
\ea
Here we have used the relation $\sqrt{s}/2 = \o_{J_c}$, valid for
both jets in the c.m.  At the same time, we make the identification,
\be
{1\over s} \delta\left({\sqrt{s}\over 2} - \o(N_{J_c})\right)\,
\delta^2(\hat n_c - \hat n(N_{J_c}))
=
{1\over 4}\, \delta^3\left(\vec p_{J_c}-\vec p(N_{J_c})\right)\, ,
\ee
which again holds in the c.m.\ frame.  The spatial components
of each $p_{J_c}$ are thus fixed.  Given that we are at small
$\bar{\varepsilon}_{J_c}$,
the jet functions may be thought of as  functions of
the light-like jet momenta $p_{J_c}^\mu$ of Eq.\ (\ref{lightlike})
and of $\bar{\varepsilon}_{J_c}$.  Because the vector jet function is
constructed
to be dimensionless, $\bar{J}'{}_c^\mu$ in Eq.\ (\ref{jetdef})
is proportional to $\beta_c$
rather than $p_{J_c}$.  Otherwise, it is free of explicit
$\beta_c$-dependence.

The jet functions can now be written in terms of boost-invariant
arguments,
homogeneous of degree zero in $\xi_c$:
\ba
\bar J_c\left(\bar{\varepsilon}_{J_c},a,\mu\right) &=&
\bar{\beta}_c{\,}_\mu \ \Bigg [\
   \beta_c^\mu
\, \bar{J}_c^{(1)} \left(\frac{p_{J_c} \cdot \xic}{\mu},
\bar{\varepsilon}_{J_c} \, \frac{\sqrt{s}}{\mu} \, \left(
\frac{\sqrt{s}}{2 p_{J_c} \cdot \xic} \right)^{1-a}, a,\as(\mu)
\right)
\nonumber
\\
&\ & \hspace{-5mm} +\  \, \frac{2\, \xi_c^\mu\ \beta_c\cdot \xi_c
}{\left|\xi_c\right|^2} \,
\bar{J}_c^{(2)} \left(\frac{p_{J_c} \cdot \xic}{\mu},
\bar{\varepsilon}_{J_c} \, \frac{\sqrt{s}}{\mu} \, \left(
\frac{\sqrt{s}}{2 p_{J_c} \cdot \xic} \right)^{1-a}, a, \as(\mu)
\right) \Bigg ]\, ,
\label{primedef2}
\ea
where ${\bar J}^{(1)}$ and ${\bar J}^{(2)}$ are independent functions, and
where we have suppressed possible dependence on
${\hat \xi}_{c, \, \perp}$.
For jet $c$, the weight $\bar{\varepsilon}_{J_c}$ is fixed by
$\delta(\bar{\varepsilon}_{J_c}-\bar{f}_c(N_{J_c},a))$,
where on the right-hand side of  the expression for the weight (\ref{fbarLC1}),
the sum over each particle's momentum involves the overall factor
$(2 p_{J_c}^\pm/\sqrt{s})^{1-a}$.
After integration over final states at fixed $\bar{\varepsilon}_{J_c}$,
the jet can thus depend on the vector $p_{J_c}^\mu$.
At the same time, it is easy to see from the definition
of the weight that $p_{J_c}^\mu$ can only appear
in the combination $(1/\bar{\varepsilon}_{J_c} \sqrt{s})^{1/(1-a)}\,
(2 p_{J_c}^\mu/\sqrt{s})$.
This vector can combine with $\xi_c$ to form an invariant, and
all $\xi_c$-dependence comes about in this way.

Expression (\ref{primedef2}) can be further simplified by noting that
\be
2\, \bar{\beta}_c \cdot  \xi_c\ {\beta}_c \cdot  \xi_c    =
\xi_c^2 + \xi_{c,\,\perp}^2\,   \, .
\ee
Choosing $\xi_{c,\,\perp} = 0$, we find a single combination,
\ba
\bar J_c\left(\bar{\varepsilon}_{J_c},a,\mu\right)
=
\bar J_c\left( \frac{p_{J_c} \cdot \xic}{\mu},
   \bar{\varepsilon}_{J_c} \, \frac{\sqrt{s}}{\mu} \, \left( \zeta_c
\right)^{1-a}, a, \as(\mu)
\right)\, ,
\ea
where, in the notation of Eq.\ (\ref{primedef2}), $\bar J_c=\bar
J_c^{(1)}+\bar J_c^{(2)}$, and we have defined
\ba
\zeta_c\equiv {\sqrt{s} \over 2 p_{J_c}\cdot \xic} \, .
\label{zetadef}
\ea
In these terms, the Laplace  moments of the jet function inherit
dependence on the
moment variable $\nu$ through
\ba
J_c \left(\nu,a,\mu \right)
&=& \int_0^\infty d\bar{\varepsilon}_{J_c} \; {\rm e}^{-\nu 
\bar{\varepsilon}_{J_c}}\, \bar
J_c\left(\bar{\varepsilon}_{J_c},a,\mu\right)
\nonumber\\
& \equiv &
J_c\left(\frac{p_{J_c} \cdot \xic}{\mu}, \frac{\sqrt{s}}{\mu \nu} \,
\left(\zeta_c \right)^{1-a},
a,\as(\mu) \right),
\label{primedef}
\ea
where the unbarred and barred quantities denote transformed and
untransformed functions, respectively. We have constructed
the jet functions to be independent of $\varepsilon$, since the radiation
into $\O$
is at wide angles from the jet axes and can therefore be completely
factored from the collinear radiation. This radiation at wide angles
is contained in the soft function, which will be defined below in
a manner that avoids double counting in the cross section.

\subsection{The soft function}

Given  the definitions for the jet functions in the
previous subsection, and the factorization (\ref{factor}),
we may in principle calculate  the soft function $S$
order by order in perturbation theory.
We can derive a more explicit definition of the soft function,
however, by relating it to an eikonal
analog of Eq.\  (\ref{factor}).

As reviewed  in  Refs.\ \cite{BKS1,pQCD},
soft radiation at wide angles from the jets decouples
from the collinear lines within the jet.
As a result, to
compute amplitudes for wide-angle radiation, the jets
may be replaced by nonabelian phases, or Wilson lines.
We therefore construct a dimensionless
quantity, $\sigma^{(\mbox{\tiny eik})}$,
in which gluons are radiated by path-ordered exponentials
$\Phi$, which mimic the color flow of outgoing quarks,
\be
\Phi^{({\mathrm f})}_{\beta_c} (\infty,0;x)  =  P e^{-i g \int_{0}^{\infty} d
\lambda \beta_c \cdot {\mathcal{A}}^{({\mathrm f})} (\lambda \beta_c + x )},
\ee
with $\beta_c$ a light-like velocity in either of the jet directions.
For the two-jet cross section at measured
$\varepsilon$ and $\bar \varepsilon_{\mbox{\tiny eik}}$, we define
\ba
\bar{\sigma}^{(\mbox{\tiny
eik})}\left(\varepsilon,\bar{\varepsilon}_{\mbox{\tiny eik}}, a ,\mu
\right)\!\!\!
&\!\! \equiv  \!\!\!& \!\!\!{1\over \Ncol}\
\sum_{N_{\mbox{\tiny eik}}} \left< 0
\left| \Phi^{(\bar {\rm q})}_{\beta_2}{}^\dagger(\infty,0;0)
\Phi^{(\rm q)}_{\beta_1}{}^\dagger(\infty,0;0)
\right| {N_{\mbox{\tiny eik}}} \right>
\nonumber \\
& \ & \hspace{5mm} \times\;
\left< N_{\mbox{\tiny eik}}
\left| \Phi^{(\rm q)}_{\beta_1}(\infty,0;0)
\Phi_{\beta_2}^{(\bar{\rm q})}(\infty,0;0)
\right| 0 \right>  \ \delta\left(\varepsilon - f(N_{\mbox{\tiny eik}})
\right)
\delta\left(\bar{\varepsilon}_{\mbox{\tiny eik}} -
\bar{f}(N_{\mbox{\tiny eik}},a) \right)
\nonumber\\
&=& \delta(\varepsilon)\,  \delta(\bar{\varepsilon}_{\mbox{\tiny
eik}}) +{\cal O}(\alpha_s)\,
.
\label{eikdef}
\ea
  The sum is over all final states $N_{\mbox{\tiny eik}}$ in
the eikonal cross section. The renormalization
scale in this cross section, which will also serve as a factorization
scale, is denoted $\mu$.  Here the event shape function
$\bar{\varepsilon}_{\mbox{\tiny eik}}$
is defined by $\bar{f}(N_{\rm eik},a)$ as in Eqs.\  (\ref{barfdef}) and
(\ref{2jetf}),
distinguishing between the hemispheres around the jets.
As usual, $\Ncol$  is the number of colors,
and a trace over color is understood.

The eikonal cross section (\ref{eikdef}) models the soft
radiation away from the jets, including the radiation into $\O$,
accurately.
It also contains enhancements
for configurations collinear to the jets, which, however,
are  already taken into account by the partonic jet
functions in (\ref{factor}).  Indeed, (\ref{eikdef}) does not reproduce
the
partonic cross section accurately for collinear radiation.
It is also easy to verify at lowest
order that even at fixed $\bar{\varepsilon}_{\mbox{\tiny eik}}$
the eikonal cross section (\ref{eikdef}) is
ultraviolet divergent in dimensional regularization,
unless we also impose a cutoff on
the energy of real gluon emission collinear to $\beta_1$
or $\beta_2$.

The construction of the soft function $S$
from $\bar{\sigma}^{(\mbox{\tiny eik})}$ is nevertheless possible
   because the eikonal cross
section (\ref{eikdef}) factorizes in the same manner
as the cross section itself, into eikonal jet functions
and a soft function.  The essential point \cite{KOS} is that
the soft function in the factorized eikonal cross section
is the same as in the original cross section (\ref{factor}).
The eikonal jets organize all collinear enhancements
in (\ref{eikdef}), including the spurious ultraviolet
divergences.  These eikonal jet functions are defined
analogously to their partonic counterparts, Eq.\ (\ref{jetdef}),
but now with ordered exponentials replacing the quark fields,
\ba
\bar{J}_c^{(\mbox{\tiny eik})}\left(\bar{\varepsilon}_c,a,\mu \right)
& \equiv  & {1\over \Ncol}\,
\sum_{N_c^{(\mbox{\tiny eik})}}
\left<0 \left| \Phi^{({\mathrm f}_c)}_{\xi_c}{}^\dagger(0,-\infty;0)
\Phi_{\beta_c}^{({\mathrm f}_c)}{}^\dagger(\infty,0;0) \right|
N_c^{(\mbox{\tiny eik})} \right>
\nonumber \\
& \ & \hspace{5mm} \left< N_c^{(\mbox{\tiny eik})} \left|
\Phi^{({\mathrm f}_c)}_{\beta_c}(\infty,0;0)
\Phi_{\xi_c}^{({\mathrm f}_c)}(0,-\infty;0) \right|  0 \right> \,
\delta\left(\bar{\varepsilon}_c -  \bar{f}_c(N_c^{(\mbox{\tiny
eik})},a) \right)
\nonumber\\
&=& \delta(\bar{\varepsilon}_c) +{\cal O}(\alpha_s)\, ,
\label{eikjetdef}
\ea
where ${\mathrm f}_c$ is a quark or antiquark, and where the
trace over color is understood.
The weight functions are given as above, by Eq.\ (\ref{fbar2jet1}),
with the sum over particles in all directions.

In terms of the eikonal jets, the eikonal cross section (\ref{eikdef})
factorizes as
\ba
\bar{\sigma}^{(\mbox{\tiny
eik})}\left(\varepsilon,\bar{\varepsilon}_{\mbox{\tiny eik}},a,\mu \right)
& \equiv  &
\int  d \bar{\varepsilon}_s \,
\bar{S}\left(\varepsilon,\bar{\varepsilon}_s,a,\mu \right)
\prod\limits_{c = 1}^2 \int d \bar{\varepsilon}_c \,
\bar{J}_c^{(\mbox{\tiny eik})}\left(\bar{\varepsilon}_c,a,\mu \right)\;
    \delta \left(\bar{\varepsilon}_{\mbox{\tiny eik}} -
\bar{\varepsilon}_s-\bar{\varepsilon}_1-\bar{\varepsilon}_2 \right),
\label{eikfact}
\ea
where we pick the factorization scale equal to the renormalization scale
$\mu$.  As for the full cross section, the convolution
in (\ref{eikfact}) is simplified by a Laplace
transformation (\ref{primedef}) with respect to
$\bar{\varepsilon}_{\mbox{\tiny eik}}$,
which allows us to solve for the soft function as
\be
S \left(\varepsilon,\nu,a,\mu\right) =
\frac{\sigma^{(\mbox{\tiny eik})}\left(\varepsilon,\nu,a,\mu \right) }
{\prod\limits_{c = 1}^2 J_c^{(\mbox{\tiny eik})}\left(\nu,a,\mu\right) }
=\delta(\varepsilon)+{\cal O}(\alpha_s)\, .
\label{s0}
\ee
In this ratio, collinear logarithms
in $\nu$ and the unphysical ultraviolet divergences and their
associated cutoff dependence cancel between the eikonal
cross section and the eikonal jets, leaving a soft
function that is entirely free of collinear enhancements.
The soft function retains $\nu$-dependence through soft
emission, which is also restricted by the weight function
$\varepsilon$.  In addition, because soft radiation
within the eikonal jets can be factored from its collinear
radiation, just as in the partonic jets, all
logarithms in $\nu$ associated with wide-angle radiation
are identical between the partonic and eikonal jets,
and factor from logarithmic corrections associated with
collinear radiation in both cases.
As a result, the inverse eikonal jet
functions cancel contributions from the wide-angle soft radiation of
the partonic jets in the
transformed cross section (\ref{trafosig}).

Given the definition
of the energy flow weight function $f$, Eq.\ (\ref{eflowdef}), the 
soft function is not
boost invariant.  In addition, because it is  free of
collinear logs, it can have at most a single
logarithm per loop.  Its dependence on $\varepsilon$
is therefore only through ratios of the dimensional
quantities $\varepsilon\sqrt{s}$  with the renormalization
(factorization) scale.

As in the case of the partonic jets, Eq.\ (\ref{primedef}),  we need to
identify
the variable through which $\nu$ appears in the soft
function.
We note that dependence on the velocity vectors $\beta_c$
and the factorization vectors $\xi_c$ must be scale invariant
in each, since they arise only from eikonal lines and vertices.
The eikonal jet functions cannot depend explicitly on the scale-less, lightlike
eikonal velocities $\beta_c$, and  $\sigma^{\rm (eik)}$
is independent of the $\xi_c$.  Dependence on the factorization
vectors $\xi_c$ enters only
   through the weight functions, (\ref{fbarLC1}) for the eikonal
jets, in a manner analogous to the case of the partonic jets. This results in
a dependence on $(\zeta_c)^{1-a}$, as above, with $\zeta_c$ defined in
Eq. (\ref{zetadef}).  In summary, we may characterize the arguments 
of the soft function in
transform space as
\be
S\left(\varepsilon,\nu,a,\mu \right) =
S\left(\frac{\varepsilon \sqrt{s}}{\mu},\varepsilon\nu,
\frac{\sqrt{s}}{\mu \nu} \, \left( \zeta_c \right)^{1-a},
a, \as(\mu)
\right)\, .
\label{Sargs}
\ee

\section{Resummation}

We may summarize the results of the previous
section by rewriting the transform of the factorized cross section
(\ref{trafosig})
in terms of the hard, jet and soft functions identified above,
which depend on the kinematic variables and the moment $\nu$
according to Eqs.\ (\ref{harddef}), (\ref{primedef}) and (\ref{Sargs})
respectively,
\ba
\frac{d \sigma \left(\varepsilon,\nu,s ,a\right)}{d\varepsilon\,d
\hat{n}_1 }
&=&
   {d \sigma_0 \over d\hat{n}_1}\ H \left(
\frac{\sqrt{s}}{\mu},\frac{p_{J_c} \cdot \xic}{\mu},\hat{n}_1,\as(\mu)
\right)\,
\prod_{c=1}^2\;
J_c\left(\frac{p_{J_c} \cdot \xic}{\mu}, \frac{\sqrt{s}}{\mu \nu} \,
(\zeta_c)^{1-a},
a,\as(\mu) \right)\,
\nonumber \\
& \ &\hspace{15mm} \times\
    S\left(\frac{\varepsilon \sqrt{s}}{\mu}, \varepsilon \nu,
\frac{\sqrt{s}}{\mu \nu}\left(\zeta_c \right)^{1-a},
a, \as(\mu)
\right)\, .
    \label{factorcom}
\ea
The natural scale for the strong coupling
in the short-distance function $H$ is $\sqrt{s}/2$.
Setting $\mu = \sqrt{s}/2$, however, introduces large logarithms of
$\varepsilon$ in the soft function and large logarithms of $\nu$ in both the
soft and jet functions.
The purpose of this section is to control these logarithms by
the identification and solution of  renormalization
group and evolution equations.

The information necessary to perform the resummations is already present in
  the factorization (\ref{factorcom}).
The cross section itself is independent of the factorization scale
\ba
\mu \frac{d}{d \mu} \frac{d \sigma \left(\varepsilon,\nu,s, a \right)}{d
\varepsilon d\hat{n}_1 }
& = & 0\, ,
\label{muev} \
\ea
   and of the choice of
the eikonal directions, $\xic$, used in the factorization,
\ba
\frac{\partial}{\partial \ln  \left(p_{J_c} \cdot \hat{\xi}_c\right) }
\frac{d \sigma \left(\varepsilon,\nu,s, a \right)}{d \varepsilon d\hat{n}_1 }
& = & 0\, .
\label{xiev}
\ea
The arguments of this section
closely follow the analysis of Ref.\ \cite{CLS}.
We will see that the dependence of
jet and soft functions on the parameter $a$
that characterizes
the event shapes (3) is reflected
in the resummed correlations, so that
the relationship between correlations with
different values of $a$ is both calculable
and nontrivial.

\subsection{Energy flow}

As a first step, we use the renormalization group equation (\ref{muev})
to organize dependence on the energy flow variable $\varepsilon$.
Applying Eq.\ (\ref{muev}) to the factorized correlation (\ref{factorcom}), we
derive the following consistency conditions, which are themselves
renormalization
group equations:
\ba
\mu \frac{d}{d \mu}\;
\ln\, S\left(\frac{\varepsilon \sqrt{s}}{\mu}, \varepsilon \nu,
\frac{\sqrt{s}}{\mu \nu} (\zeta_c)^{1-a},
a, \as(\mu) \right)
& = & -
\gamma_s\left(\as(\mu)\right),
\label{softmu}
\\
\mu \frac{d}{d \mu}\;
\ln\, J_c\left(\frac{ p_{J_c} \cdot \xic }{\mu },
   \frac{\sqrt{s}}{\mu \nu} \, (\zeta_c)^{1-a} ,
a,\as(\mu) \right) & = & - \gamma_{J_c}\left(\as(\mu)\right),
\label{jetmu}
\\
\mu \frac{d}{d \mu}\; \ln\,  H\left( \frac{\sqrt{s}}{\mu},\frac{p_{J_c} \cdot
\xic}{\mu},\hat{n}_1,\as(\mu) \right)
&=& \gamma_s\left(\alpha_s(\mu)\right)
+ \sum_{c=1}^2\gamma_{J_c}\left(\alpha_s(\mu)\right)\, .
\label{Hmu}
\ea
The anomalous dimensions $\gamma_d$, $d=s,\, J_c$ can
depend only on variables held in common between  at least two
of the functions.  Because each function is infrared safe,
while ultraviolet divergences are present only in virtual
diagrams, the anomalous dimensions cannot depend on
the parameters $\nu$, $\varepsilon$ or $a$.  This leaves
as arguments of the $\gamma_d$ only
the  coupling $\as(\mu)$, which we exhibit, and $\zeta_c$, which
we suppress for now.

Solving Eqs. (\ref{softmu})
and (\ref{jetmu}) we find
\ba
S\left(\frac{\varepsilon \sqrt{s}}{\mu}, \varepsilon \nu,
\frac{\sqrt{s}}{\mu \nu} \left(\zeta_c \right)^{1-a},
a, \as(\mu) \right)
& = &
S\left(\frac{\varepsilon \sqrt{s}}{\mu_0}, \varepsilon \nu,
\frac{\sqrt{s}}{\mu_0 \nu} \left( \zeta_c \right)^{1-a},
a, \as(\mu_0) \right) \, e^{-\int\limits_{\mu_0}^\mu \frac{d
\lambda}{\lambda} \gamma_s\left(\as(\lambda)\right)},
\label{softevol}
\nonumber \\
&\ & \\
J_c \left(\frac{ p_{J_c} \cdot \xic }{\mu},
   \frac{\sqrt{s}}{\mu \nu} \, \left(\zeta_c \right)^{1-a} , a, \as(\mu)
\right)
& = &
J_c \left( \frac{p_{J_c} \cdot \xic }{\tilde{\mu}_0 },
\frac{\sqrt{s}}{\tilde{\mu}_0 \nu} \, \left(\zeta_c \right)^{1-a} ,
a,\as(\tilde{\mu}_0) \right)
    \,e^{-\int\limits_{\tilde{\mu}_0}^\mu \frac{d \lambda}{\lambda}
\gamma_{J_c}\left(\as(\lambda)\right)}\, , \nonumber
\\
&\ &\label{jetevol}
\ea
for the soft and jet functions.  As suggested above, we will eventually pick
$\mu\sim \sqrt{s}$
to avoid large logs in $H$.
Using these expressions in Eq. (\ref{factorcom}) we can avoid
logarithms of $\varepsilon$ or $\nu$ in the soft function, by evolving from
$\mu_0 = \varepsilon \sqrt{s}$ to the factorization scale $\mu \sim \sqrt{s}$.
No choice of $\tilde{\mu}_0$, however, controls all logarithms of $\nu$ in
the jet functions.  Leaving $\tilde \mu_0$ free, we find for the
cross section (\ref{factorcom}) the
intermediate result
\ba
    \label{resume}
\frac{d \sigma \left(\varepsilon, \nu,s ,a\right)}{d\varepsilon \, d
\hat{n}_1 }
&=&
{d \sigma_0 \over d\hat{n}_1}\ H \left( \frac{\sqrt{s}}{\mu},\frac{p_{J_c}
\cdot \xic}{\mu},\hat{n}_1,\as(\mu) \right)\,
\nonumber\\
&\ & \hspace{-15mm} \times\; S\left(1, \varepsilon \nu, (\zeta_c)^{1-a}, a
,\as(\varepsilon \sqrt{s})
\right)\,
\exp\left\{ -\int\limits_{\varepsilon \sqrt{s}}^{\mu} \frac{d
\lambda}{\lambda} \, \gamma_s\left(\as(\lambda)\right)\right\}
\\
&\ & \hspace{-15mm} \times\;
J_c \left( \frac{p_{J_c} \cdot \xic }{\tilde{\mu}_0 },
\frac{\sqrt{s}}{\tilde{\mu}_0 \nu} \, \left(\zeta_c \right)^{1-a} ,
a,\as(\tilde{\mu}_0) \right)
    \,\exp\left\{-\int\limits_{\tilde{\mu}_0}^\mu \frac{d \lambda}{\lambda}
\gamma_{J_c}\left(\as(\lambda)\right)\right\} \, . \nonumber
\ea
We have avoided introducing logarithms of $\varepsilon$ into the jet functions,
which originally only depend on $\nu$, by evolving the soft and the
jet functions independently.
The choice of $\mu_0=\varepsilon\sqrt{s}$ or
$\sqrt{s}/\nu$ for the soft function is to some extent a matter of
convenience,
since the two choices differ by logarithms of $\varepsilon\nu$.
In general, if we choose $\mu_0=\sqrt{s}/\nu$, logarithms
of $\varepsilon\nu$ will appear multiplied by coefficients that reflect
the
size of region $\O$.  An example is Eq.\ (\ref{ps0}) above.
When $\O$ has a small angular size, $\mu_0=\sqrt{s}/\nu$ is generally the
more natural choice, since then logarithms in $\varepsilon\nu$ will
enter with small weights.  In contrast, when $\O$ grows to cover most
angular directions, as in
the study of rapidity gaps \cite{Oderda}, it is more
natural to choose $\mu_0 = \varepsilon\sqrt{s}$.

\subsection{Event shape transform}

The remaining unorganized ``large" logarithms in (\ref{resume}),
are in the jet functions,
which we will resum by using the consistency equation (\ref{xiev}).
The requirement that the cross section be independent of $p_{J_c}\cdot
\hat{\xi}_c$
implies that the jet, soft and hard functions obey equations
analogous to (\ref{softmu})--(\ref{Hmu}), again in terms of the variables
that they hold in common \cite{CLS}.  The same results may
be derived following the method of Collins and Soper
   \cite{ColSop}, by defining the jets in an axial gauge,
and then studying their variations under boosts.

For our purposes, only the equation satisfied by the
jet functions \cite{ColSop,CLS} is necessary,
\ba
    \frac{\partial }{\partial \ln \left(p_{J_c} \cdot \xic\right)}
\ln\ J_c \left( \frac{p_{J_c} \cdot \xic}{\mu},
\frac{\sqrt{s}}{\mu \nu} \, (\zeta_c)^{1-a} ,a,\as(\mu)
\right)
& \ & \nonumber
\\
&\ & \hspace{-70mm} =
    K_c\left(\frac{\sqrt{s}}{\mu\,
\nu}(\zeta_c)^{1-a},
a,\as(\mu)
\right)   +  G_c\left(\frac{p_{J_c} \cdot \xic}{\mu},\as(\mu)\right)   \, .
    \label{KGend}
\ea
The functions $K_c$ and $G_c$
compensate
the $\xi_c$-dependence of the soft and hard functions, respectively,
which determines the kinematic variables upon which they may depend.
In particular, notice the combination of $\nu$- and $\xi_c$-dependence
required by the arguments of the jet function, Eq.\ (\ref{primedef}).

Since the definition of our jet functions (\ref{jetdef}) is gauge
invariant,
we can derive the kernels $K_c$ and $G_c$ by an explicit  computation of
   ${\partial\, J_c}/{\partial \ln
\left(p_{J_c} \cdot \hat{\xi}_c\right)}$ in any gauge.
The multiplicative renormalizability of the jet function, Eq. (\ref{jetmu}),
with an anomalous dimension that is independent of $p_{J_c}\cdot \xic$
ensures that the right-hand side of Eq.\ (\ref{KGend}) is
a renormalization-group invariant.  Thus, $K_c+G_c$ are renormalized
additively, and satisfy \cite{ColSop}
\ba
\mu \frac{d}{d \mu}\
K_c\left(\frac{\sqrt{s}}{\mu\, \nu}\left(\zeta_c\right)^{1-a},
a,\as(\mu) \right) & = & - \gamma_{K_c}
\left(\as(\mu)\right),
\nonumber\\
\mu \frac{d}{d \mu}G_c\left(\frac{p_{J_c} \cdot \xic }{\mu},\as(\mu)\right)
   & = &  \gamma_{K_c}
\left(\as(\mu)\right) \, .
\label{Gevol}
\ea
Since $G_c$ and hence
$\gamma_{K_c}$, may be computed from
virtual diagrams, they do not depend on $a$, and $\gamma_{K_c}$ is the
universal Sudakov anomalous dimension \cite{ColSop,sudgam}.

With the help of these evolution equations, the terms $K_c$ and $G_c$
in Eq. (\ref{KGend}) can be reexpressed as
\cite{cssdy}
\ba
    K_c\left(\frac{\sqrt{s}}{\mu\, \nu}\left(\zeta_c\right)^{1-a},a,\as(\mu)
\right)   +  G_c\left(\frac{p_{J_c} \cdot \xic}{\mu},\as(\mu)\right)
&\ &  \nonumber \\
&\ & \hspace{-90mm} =
K_c\left(\frac{1}{c_1},a,\as\left(c_1 \,
\frac{\sqrt{s}}{\nu}\left(\zeta_c\right)^{1-a}\right)
\right)
+  G_c\left(\frac{1}{c_2},\as\left(c_2 \, p_{J_c} \cdot \xic \right) \right)
\ - \int\limits_{c_1 {\sqrt{s}}\, \left(\zeta_c\right)^{1-a}/{\nu} }^{ c_2\,
p_{J_c} \cdot \xic }
\frac{d  \lambda'}{\lambda'} \gamma_{K_c}\left(\as\left(\lambda'\right)
\right)
\nonumber \\
&\ & \hspace{-90mm} =
    - B'_c\left(c_1,c_2,a,  \as\left(c_2 \, p_{J_c} \cdot \xic \right) \right)
-
2 \int\limits_{c_1 {\sqrt{s}}\, \left(\zeta_c\right)^{1-a}/{\nu} }^{ c_2 \,
p_{J_c}
\cdot \xic }
\frac{d  \lambda'}{\lambda'} A'_c\left(c_1, a, \as\left(\lambda'\right)
\right)\, ,
\label{ABabbr}
\ea
where in the second equality we have shifted the argument of
the running coupling in $K_c$, and have introduced the notation
\ba
B'_c\left(c_1,c_2,a, \as\left(\mu \right) \right)
& \equiv & -
K_c\left(\frac{1}{c_1},a, \as\left(\mu \right)  \right) -
G_c\left(\frac{1}{c_2} ,\as\left(\mu \right)\right),
\nonumber \\
2 A'_c\left( c_1, a, \as\left(\mu \right) \right) & \equiv &  \gamma_{K_c}
\left(\as(\mu) \right) + \beta(g(\mu)) \frac{\partial}{\partial
g(\mu)} K_c\left(\frac{1}{c_1},a, \as(\mu)\right).
\label{ABdef}
\ea
The primes on the functions $A'_c$ and $B'_c$ are to distinguish
these anomalous dimensions from their somewhat more familiar versions
given below.

The solution to Eq. (\ref{KGend}) with $\mu = \tilde{\mu}_0$ is
\ba
J_c \left( \frac{p_{J_c} \cdot \xic }{\tilde{\mu}_0},
   \frac{\sqrt{s}}{\tilde{\mu}_0 \nu} \, \left(\zeta_c
   \right)^{1-a} ,a,\as(\tilde{\mu}_0)
\right)
&=&
J_c \left( \frac{\sqrt{s}}{2 \,\zeta_0 \,\tilde{\mu}_0},
   \frac{\sqrt{s}}{\tilde{\mu}_0 \nu} \, \left(\zeta_0
   \right)^{1-a} ,a,\as(\tilde{\mu}_0)
\right)   \nonumber \\
&\ & \hspace{-67mm}
    \times \, \exp \left\{ -\int\limits_{\sqrt{s}/(2\zeta_0) }^{p_{J_c} \cdot
\xic}
    \frac{d \lambda}{\lambda} \left[B'_c\left(c_1,c_2,a,
\as\left(c_2 \lambda \right) \right)  +  2 \int\limits_{c_1 \frac{s^{1-a/2}
}{ \nu
(2\,\lambda)^{1-a} } }^{c_2\, \lambda}\frac{d \lambda'}{\lambda'} A'_c\left(
c_1,
a,\as\left(\lambda'\right) \right) \right] \right\}\, ,
   \label{orgsol}
\ea
where we evolve from $\sqrt{s}/(2\,\zeta_0)$ to $p_{J_c} \cdot \xic =
\sqrt{s}/(2 \,\zeta_c)$ (see Eq.\ (\ref{zetadef})) with
\be
\zeta_0 = \left(\frac{\nu}{2}\right)^{1/(2-a)}. \label{zeta0}
\ee
After combining Eqs.\ (\ref{jetevol}) and (\ref{orgsol}),
the choice $\tilde{\mu}_0 = \sqrt{s}/(2\zeta_0) = \frac{\sqrt{s}}{\nu}
(\zeta_0)^{1-a}$
   allows us to control
all large logarithms in
the jet functions simultaneously:
\footnote{After this paper was submitted for publication,
a related analysis of event shape and
energy flow correlations was given by Dokshitzer and
Marchesini \cite{DM03}, who identify the same
factorization of soft radiation described here
and in \cite{BKSproc}, and who
study the leading logarithms of $\varepsilon\nu$
for $\varepsilon\ll 1/\nu$, using the methods of
\cite{BanfiMarchSmye}.}
\ba
J_c \left( \frac{p_{J_c} \cdot \xic}{\mu},
\frac{\sqrt{s}}{\mu \nu} \, (\zeta_c)^{1-a} ,a,\as(\mu)
\right)
&=&
J_c \left(
1, 1,a,\as\left(\frac{\sqrt{s}}{2 \, \zeta_0} \right) \right)
\,\exp \left\{-\int\limits_{\sqrt{s}/(2 \zeta_0)}^\mu
\frac{d\lambda}{\lambda} \gamma_{J_c} \left(\as(\lambda)\right) \right\}
\, \nonumber \\
&\ & \hspace{-60mm}
    \times \, \exp \left\{ -\int\limits_{\sqrt{s}/(2\, \zeta_0) }^{p_{J_c}
\cdot \xic}
    \frac{d \lambda}{\lambda} \left[B'_c\left(c_1,c_2,a, \as\left(c_2 \lambda
\right) \right) + 2 \int\limits_{c_1 \frac{s^{1-a/2} }{ \nu
(2\,\lambda)^{1-a} } }^{c_2\, \lambda}\frac{d \lambda'}{\lambda'} A'_c\left(
c_1,
a,\as\left(\lambda'\right) \right) \right] \right\}\, .
   \label{jetxiend}
\ea
As observed above, we treat $a$ as a fixed parameter, with $|a|$ 
small compared to
$\ln\, (1/\varepsilon)$ and $\ln\nu$.

\subsection{The resummed correlation}

Using Eq. (\ref{jetxiend}) in (\ref{resume}), and setting $\mu = \sqrt{s}/2$,
we find a fully resummed form for the correlation,
\ba
\frac{d \sigma \left(\varepsilon, \nu,s,a \right)}{d \varepsilon\, d
\hat{n}_1 }
&=&
{d \sigma_0 \over d\hat{n}_1}\ H \left(\frac{2 \, p_{J_c}
\cdot \xic}{\sqrt{s}},\hat{n}_1,\as\left(\frac{\sqrt{s}}{2}\right) \right)\,
   \nonumber \\
& \ & \hspace*{-2cm}
\times \, S\left(1, \varepsilon \nu, (\zeta_c)^{1-a}, a,\as(\varepsilon
\sqrt{s} ) \right)
\, \exp \left\{  - \int\limits_{\varepsilon \sqrt{s}}^{\sqrt{s}/2} \frac{d
\lambda}{\lambda} \gamma_s\left(\as(\lambda)\right) \right\} \nonumber \\
& \ & \hspace*{-2cm}
\times \, \prod_{c=1}^2\, J_c \left(1,1,a,\as\left(\frac{\sqrt{s}}{2 \,
\zeta_0}\right) \right)
    \exp \left\{ - \int\limits_{\sqrt{s}/(2 \, \zeta_0)}^{\sqrt{s}/2}
   \frac{d \lambda}{\lambda} \gamma_{J_c} \left(\as(\lambda)\right) \right\}
\nonumber \\
& \ & \hspace*{-2cm}
\times \, \exp \left\{ -\int\limits_{\sqrt{s}/(2\, \zeta_0) }^{p_{J_c} \cdot
\xic}
    \frac{d \lambda}{\lambda} \left[B'_c\left(c_1,c_2,a,
\as\left(c_2 \lambda \right) \right)  +  2 \int\limits_{c_1 \frac{s^{1-a/2}
}{ \nu
(2\,\lambda)^{1-a} } }^{c_2\, \lambda}\frac{d \lambda'}{\lambda'} A'_c\left(
c_1,
a,\as\left(\lambda'\right) \right) \right] \right\}\, . \nonumber \\
& & \label{evolend}
\ea

Alternatively, we can combine all jet-related exponents in Eq.
(\ref{evolend}) in the correlation.
As we will verify below in Section
\ref{gauge}, the cross section is independent of the choice of $\xi_c$.
As a result, we can choose
\be
p_{J_c} \cdot \xic = \frac{\sqrt{s}}{2}\, .
\label{xiid}
\ee
This choice allows us to combine $\gamma_{J_c}$ and
$B'_c$ in Eq. (\ref{evolend}),
\ba
\frac{d \sigma \left(\varepsilon, \nu,s,a \right)}{d \varepsilon\, d
\hat{n}_1 }
&=&
{d \sigma_0 \over d\hat{n}_1}\ H \left(1,
\hat{n}_1,\as\left(\frac{\sqrt{s}}{2}\right) \right)\,
   \nonumber \\
& \ & \hspace*{-3cm}
\times \, S\left(1, \varepsilon \nu, 1, a,\as(\varepsilon \sqrt{s} ) \right)
\, \exp \left\{  - \int\limits_{\varepsilon \sqrt{s}}^{\sqrt{s}/2} \frac{d
\lambda}{\lambda} \gamma_s\left(\as(\lambda)\right) \right\} \,
\prod_{c=1}^2\, J_c \left(1,1,a,\as\left(\frac{\sqrt{s}}{2 \, \zeta_0}\right)
\right) \nonumber \\
& \ & \hspace*{-3cm}
\times \, \exp \left\{ -\int\limits_{\sqrt{s}/(2\, \zeta_0) }^{\sqrt{s}/2}
    \frac{d \lambda}{\lambda} \left[ \gamma_{J_c} \left(\as(\lambda)\right) +
B'_c\left(c_1,c_2,a,
\as\left(c_2 \lambda \right) \right)  +  2 \int\limits_{c_1 \frac{s^{1-a/2}
}{ \nu
(2\,\lambda)^{1-a} } }^{c_2\, \lambda}\frac{d \lambda'}{\lambda'} A'_c\left(
c_1,
a,\as\left(\lambda'\right) \right) \right] \right\}\, , \nonumber \\
& & \label{evolendnoxi}
\ea
with $\zeta_0$ given by Eq. (\ref{zeta0}).

In Eqs. (\ref{evolend}) and (\ref{evolendnoxi}), the energy flow
$\varepsilon$ appears at the level
of one logarithm per loop, in $S$, in the first exponent.
Leading logarithms of $\varepsilon$ are
therefore resummed by knowledge of $\gamma_s^{(1)}$,
the one-loop soft anomalous dimension, where we employ the
standard notation,
\ba
\gamma_s(\as) = \sum_{n=0}^\infty \gamma_s^{(n)}\ \left({\as\over
\pi}\right)^n
\ea
for any expansion in $\as$.
At the same time, $\nu$ appears in up to two logarithms per loop,
characteristic of conventional Sudakov resummation.  To control
$\nu$-dependence at the same level as $\varepsilon$-dependence, it is
natural to work to next-to-leading logarithm in $\nu$,
by which we mean the level $\as^k\, \ln^k\nu$ in the exponent.  As usual, this
requires one loop in
$B'_c$ and $\gamma_{J_c}$, and two loops in the
Sudakov anomalous dimension $A'_c$, Eq.\ (\ref{ABdef}).
These functions are straightforward to calculate from their definitions
given in the previous sections. Only the soft
function $S$ in Eqs.\ (\ref{evolend}) and (\ref{evolendnoxi})
contains information on the geometry of $\O$. The exponents are
partially process-dependent, but geometry-independent.  In
Section \ref{sec:res}, we will derive
explicit expressions for these quantities, suitable for
resummation to leading logarithm in $\varepsilon$ and next-to-leading
logarithm in $\nu$.

\subsection{The inclusive event shape}

It is also of interest
to consider the cross section for $e^+e^-$-annihilation into two jets
without fixing the energy of radiation into $\O$, but with the final state
radiation into all of phase space weighted according to Eq.\
(\ref{2jetf}), schematically
\be
e^+ + e^- \rightarrow J_1(p_{J_1}, \bar{f}_{\bar{\O}_1}) + J_2(p_{J_2},
\bar{f}_{\bar{\O}_2})\, ,
\ee
where $\bar{\O}_1$ and  $\bar{\O}_2$ cover the entire sphere.
This cross section can be factorized and resummed in a completely
analogous manner. The
final state is a convolution in the contributions
of the jet and soft functions to $\bar{\varepsilon}$ as in Eq.
(\ref{factor}),
but with no separate restriction on energy flow into $\O$.
All particles contribute to the event shape.
We obtain an expression very analogous to
Eq.\ (\ref{evolend}) for this inclusive event shape in transform space,
which can be written in terms of the same jet functions
as before, and a new function $S^{\rm incl}$ for soft radiation as:
\ba
\frac{d \sigma^{\rm incl} \left(\nu,s,a \right)}{ d \hat{n}_1 }
&=&
{d \sigma_0 \over d\hat{n}_1}\ H \left(\frac{2 p_{J_c}
\cdot \xic}{\sqrt{s} },\hat{n}_1,\as(\sqrt{s}/2) \right)\,
\nonumber\\
&\  &  \hspace*{-2cm} \times\  S^{\rm incl}\left((\zeta_c)^{1-a},
a,\as\left(\frac{\sqrt{s}}{\nu}
\right) \right) \,
\exp \left\{  - \int\limits_{\sqrt{s} / \nu }^{\sqrt{s}/2} \frac{d
\lambda}{\lambda} \gamma_s\left(\as(\lambda)\right) \right\} \nonumber \\
&\  &  \hspace*{-2cm} \times\ \prod_{c=1}^2\, J_c
\left(1,1,a,\as\left(\frac{\sqrt{s}}{2 \, \zeta_0} \right) \right)
    \exp \left\{ - \int\limits_{\sqrt{s} / (2 \, \zeta_0)}^{\sqrt{s}/2}
\frac{d \lambda}{\lambda}  \gamma_{J_c}\left(\as(\lambda)\right) \right\}
\nonumber \\
& \ & \hspace*{-2cm}
\times \, \exp \left\{ -\int\limits_{\sqrt{s}/(2\, \zeta_0) }^{p_{J_c} \cdot
\xic}
    \frac{d \lambda}{\lambda} \left[B'_c\left(c_1,c_2,a,
\as\left(c_2 \lambda \right) \right)  +  2 \int\limits_{c_1 \frac{s^{1-a/2}
}{ \nu
(2\,\lambda)^{1-a} } }^{c_2\, \lambda}\frac{d \lambda'}{\lambda'} A'_c\left(
c_1,
a,\as\left(\lambda'\right) \right) \right] \right\}\, . \nonumber \\
    \label{globalend}
\ea
Here the soft function $S^{\rm incl}=1 + {\cal O}(\alpha_s)$.
The double-logarithmic dependence of the shape transform is
identical to our resummed correlation, Eq.\ (\ref{evolend}). We will
show below, in Sec. \ref{inclusiveNLL}, that Eq. (\ref{globalend}) coincides
at NLL with the known result  for the thrust \cite{thrustresum}
when we  choose $a = 0$.

\section{Results at NLL}
\label{sec:res}

\subsection{Lowest order functions and  anomalous dimensions}

In this section, we describe the low-order calculations and results that
provide explicit expressions for the resummed shape/flow correlations and
inclusive event shape distributions at next-to-leading
logarithm in $\nu$ and leading logarithm in $\varepsilon$
(we refer to this level collectively as NLL below).   We go on to 
verify that for
the case $a=0$ we rederive the known result for the resummed
thrust at  NLL, and we exhibit the expressions for the
correlation that we will evaluate in Sec.\ \ref{numerics}.

\subsubsection{The soft function}

The one-loop soft anomalous dimension
is readily calculated in Feynman gauge from
the combination of virtual  diagrams in $\sigma^{\rm (eik)}$, Eq. 
(\ref{eikdef}), and
$J^{\rm  (eik)}$, Eq. (\ref{eikjetdef}),
in  Eq.\ (\ref{s0}).  The calculation and the result are equivalent
to those of Ref.\ \cite{KOS}, where the soft function was
formulated in axial gauge,
\be
\gamma_s^{(1)}  = - 2 \, C_F
\left[ \sum_{c=1}^2 \ln \left(\beta_c \cdot \xic \right) - \ln \left(
\frac{\beta_1 \cdot \beta_2}{2} \right) - 1 \right]\, .
\label{softad}
\ee
The first, $\xi_c$-dependent logarithmic term is associated with the eikonal
jets, while the second is a finite remainder from the
combination of $\sigma^{\rm (eik)}$ and $J^{\rm  (eik)}$ in (\ref{s0}).
Whenever $\xi_{c,\,\perp}=0$, the logarithmic terms cancel
identically,  leaving only the final  term, which comes from
the $\xic$ eikonal self-energy diagrams in the eikonal
jet functions.

The soft function is normalized to $S^{(0)}(\varepsilon) =
\delta(\varepsilon)$
as can be seen from (\ref{s0}).
For non-zero $\varepsilon$,
$d\sigma /d \varepsilon$ is given at lowest order  by
\be
S^{(1)} \left( \varepsilon \neq 0, \Omega\right) =
C_F \frac{1}{\varepsilon}
\int\limits_\O  d \mbox{PS}_2\,
\frac{1}{2 \pi} \frac{\beta_1 \cdot \beta_2}{\beta_1
\cdot \hat{k} \,\beta_2 \cdot \hat{k} }\, ,
\label{oneLoopSoft}
\ee
where $\mbox{PS}_2$ denotes the two-dimensional angular phase space
to be integrated over region $\O$, and $\hat k \equiv k / \omega_k$.
We emphasize again that the soft function contains the only
geometry-dependence of the
cross section. Also, $S^{(1)}$ for $\varepsilon \neq 0$ is independent of
$\nu$
and $a$.

As an example, consider a cone with opening angle $2 \delta$,
centered at angle $\alpha$ from jet 1.  In this case,
the lowest-order soft function is given by
\be
S^{(1)} \left( \varepsilon \neq 0, \alpha, \delta \right) =
C_F \frac{1}{\varepsilon}
\ln \left(\frac{1-\cos^2 \alpha }{\cos^2 \alpha - \cos^2 \delta}\right).
\label{softcone}
\ee
Similarly, we may choose $\O$ as a ring
extending angle $\delta_1$ to the right and $\delta_2$ to
the left of the plane perpendicular to the jet directions
in the center-of-mass.  In this case, we obtain
\be
S^{(1)} \left( \varepsilon \neq 0, \delta_1, \delta_2 \right) =
    C_F \frac{1}{\varepsilon}
\ln \left(
\frac{(1+ \sin \delta_1)}{(1-\sin \delta_1)}\frac{(1+ \sin
\delta_2)}{(1-\sin \delta_2)} \right)
= C_F \frac{2}{\varepsilon}\, \Delta\eta\, ,
\label{deltaeta2}
\ee
with $\Delta\eta$ the rapidity spanned by the ring.
For a ring centered around the center-of-mass ($\delta_1 =
\delta_2 = \delta$) the angular integral reduces to the form that we
encountered in the example  of Sec.\ \ref{sec:loe},
and that we will use in our
numerical examples of Sec.\ \ref{numerics}, with $\Delta\eta$
given by Eq.\ (\ref{deltaeta1}).

\subsubsection{The jet functions}

Recall from Eq. (\ref{Jnorm}) that the lowest-order jet function is given
by $J_c^{(0)} = 1.$

The anomalous dimensions of the jet functions are found to be
\be
\gamma_{J_c}^{(1)} = - \frac{3}{2} \, C_F
\, ,
\label{jetAD}
\ee
the same for each of the jets.
The jet anomalous dimensions
are process-independent, but of course flavor-dependent. The same
anomalous dimensions for final-state quark jets appear in three- and
higher-jet cross sections.

\subsubsection{The $K$-$G$-decomposition}

The anomalous dimension for the $K$-$G$-decomposition
is, as noted above, the Sudakov anomalous dimension,
\ba
\gamma_{K_c}^{(1)} 
& = &   2 \, C_F
, \\
\gamma_{K_c}^{(2)} 
& = &   K \, C_F
, \ea
also independent of the jet-direction. The well-known coefficient $K$
(not to be confused with the functions $K_c$) is given by \cite{KT}
\be
K = \left( \frac{67}{18} - \frac{\pi^2}{6} \right) C_A - \frac{10}{9} T_F
N_f,
\ee
with the normalization $T_F = 1/2$ and $N_f$ the number of quark
flavors.

$K_c$ and $G_c$, the functions that describe the evolution
of the jet functions in Eq.\ (\ref{KGend}), are given at one loop by
\ba
K_c^{(1)} \left(\frac{s^{1-a/2}}{\mu \nu} \left(2 p_{J_c} \cdot
\xic\right)^{a-1}\!\!\!\!\!\!,a\right) & = & - C_F
\, \ln\left(e^{2 \gamma_E-(1-a)}  \frac{\mu^2
\nu^2}{s^{2-a}} \left(2 p_{J_c} \cdot \xic \right)^{2(1-a)} \right) , \\
G_c^{(1)} \left(\frac{p_{J_c} \cdot \xic}{\mu} \right) & = & - C_F
\, \ln\left( e^{-1} \frac{\left(2 \, p_{J_c} \cdot \xic \right)^2}{\mu^2}
\right)\, .
\ea
Evolving them to the values of $\mu$ with which they
appear in the functions $A_c'$ and $B'_c$, Eq.\ (\ref{ABdef}),
they become
\ba
K_c^{(1)} \left(\frac{1}{c_1},a\right) & = & - C_F
\, \ln \left( e^{2 \gamma_E-(1-a)} c_1^2 \right) , \\
G_c^{(1)} \left(\frac{1}{c_2}\right) & = & - C_F
\ln \left( e^{-1} \frac{4}{c_2^2}
\right) .
\ea
Recall that  $G_c$ is computed from virtual diagrams
only, and  thus does not
depend on the weight function.  It therefore agrees with
the result found in \cite{ColSop}.
The soft-gluon contribution, $K_c$, which involves
real gluon diagrams, does depend on the cross section
being resummed.

With the definitions  (\ref{ABdef}) of $A'_c$ and
$B'_c$ we obtain
\ba
A_c^{\prime \,(1)}  & = & C_F  \label{A1prime}
, \\
A_c^{\prime\, (2)} \left(c_1,a\right) & = & \frac{1}{2} C_F
\left[ K + \frac{\beta_0}{2}
\ln \left( e^{2 \gamma_E -1 +a } c_1^2 \right) \right], \\
B_c^{\prime\, (1)} \left(c_1,c_2,a\right) & = & 2 C_F
\ln \left(
e^{ \gamma_E -1 +a/2 } \frac{2 \,c_1}{c_2} \right).
\ea
Here $\beta_0$ is the one-loop coefficient of the QCD beta-function,
$\beta_0 = \frac{1}{3} \left( 11 \Ncol - 4 T_F N_f \right)$ ($\beta(g)
= - g \frac{\as}{4 \pi} \beta_0 + \mathcal{O}(g^3)$).

\subsubsection{The hard scattering, and the Born cross section}

At NLL only the lowest-order hard scattering function contributes, which
is normalized to
\be
H^{(0)}(\alpha_s(\sqrt{s}/2)) = 1\,.
\ee
At this order the hard function is independent of the eikonal vectors
$\xi_c$, although it acquires $\xi_c$-dependence at higher order
through the factorization described in Sec.\ \ref{sec:sdf}.
For completeness,  we also
give the electromagnetic Born cross section $\frac{d\sigma_0}{d \hat n_1}$,
at fixed polar and azimuthal angle:
\be
\frac{d \sigma_0}{d \hat n_1} =
\Ncol \left( \sum_{\rm f} Q_{\rm f}^2 \right) \frac{\alpha_{\rm em}^2}{4 s}
\left( 1 + \cos^2 \theta \right),
\label{bornCross}
\ee
where $\theta$ is the c.m.\ polar angle of $\hat{n}_1$,
$e \, Q_{\rm f}$ is the charge of quark flavor $\rm f$, and 
$\alpha_{\rm em} = e^2/(4 \pi)$
is the fine
structure constant.

\subsection{Checking the $\xi_c$-dependence} \label{gauge}

It is instructive to verify how dependence on
the eikonal vectors $\xi_c$ cancels in the exponents of the
resummed cross section (\ref{evolend}) at the
accuracy at which we work, single logarithms of
$\varepsilon$, and single and double logarithms of $\nu$.
In these exponents,
$\xi_c$-dependence
enters only through the
combinations
$(\beta_c \cdot \xic)$ and
$(p_{J_c} \cdot \xic)$.

Let us introduce the following notation for the exponents in Eq.
(\ref{evolend}), to
which we will return below:
\ba
E_1 & \equiv &  - \int\limits_{\varepsilon \sqrt{s}}^{\sqrt{s}/2} \frac{d
\lambda}{\lambda} \gamma_s\left(\as(\lambda)\right) - \sum_{c=1}^2
\int\limits_{\sqrt{s}/(2 \, \zeta_0)}^{\sqrt{s}/2}
   \frac{d \lambda}{\lambda} \gamma_{J_c} \left(\as(\lambda)\right),
   \label{E1} \\
E_2 &  \equiv &  - \sum_{c=1}^2 \int\limits_{\sqrt{s}/(2\, \zeta_0) 
}^{p_{J_c} \cdot \xic}
    \frac{d \lambda}{\lambda} \left[B'_c\left(c_1,c_2,a,
\as\left(c_2 \lambda \right) \right)  +  2 \int\limits_{c_1 \frac{s^{1-a/2}
}{ \nu
(2\,\lambda)^{1-a} } }^{c_2\, \lambda}\frac{d \lambda'}{\lambda'} A'_c\left(
c_1,
a,\as\left(\lambda'\right) \right) \right]. \label{E2}
\ea
At NLL, explicit $\xi_c$ dependence is
found only in $\gamma_s$, Eq. (\ref{softad}), for $E_1$,
and in the upper limit of the $\lambda$ integral of $E_2$.
We then find that
\ba
{\partial \over  \partial \ln\beta_c\cdot\xic}\left(E_1+E_2\right)
=
2C_F\, \, \int\limits_{\varepsilon \sqrt{s}}^{\sqrt{s}/2}
\frac{d
\lambda}{\lambda}\; \frac{\as(\lambda)}{\pi}
-2C_F \int_{c_1{s^{1-a/2}\over
\nu(2p_{J_c}\cdot \xic)^{1-a}}}^{c_2\,p_{J_c}\cdot\xic}
{d\lambda'\over\lambda'}\;
\frac{\as(\lambda')}{\pi} +{\rm NNLL}
\, .
\label{gaugevar}
\ea
Here the second term stems entirely from $A^{\prime\,(1)}$, Eq.
(\ref{A1prime});
other contributions of $E_2$ are subleading.
The $\xi_c$-dependence in the exponents begins only at the level that
we do not resum, at $\as\ln(1/\varepsilon \nu)$,
which is compensated by corrections in $S(\varepsilon\nu,\as)$.
The remaining contributions are of NNLL order,
that is, proportional to $\as^k(\sqrt{s})
   \ln^{k-1} \left (\nu \, \beta_c \cdot \xic \right)$,
as may be verified by expanding the running couplings.
Thus, as required by the factorization procedure,
the relevant $\xi_c$-dependence cancels between
the resummed soft and jet functions, which give rise
to the first and second integrals, respectively, in Eq.\ (\ref{gaugevar}).

\subsection{The inclusive event shape at NLL}\label{inclusiveNLL}

We can simplify the differential event shape, Eq.\ (\ref{globalend}),
by absorbing the soft anomalous dimension $\gamma_s$ into
the remaining terms.  We will find a form that can be compared
directly to the classic  NLL  resummation for the thrust
($a=0$).   This is done by rewriting the integral over
the soft anomalous dimension as
\ba
\int_{\sqrt{s}/\nu}^{\sqrt{s}/2} {d\lambda\over \lambda}\;
\gamma_s\left(\as(\lambda)\right)
&=&
\int_{\sqrt{s}/\left[2(\nu/2)^{1/(2-a)}\right]}^{\sqrt{s}/2}
{d\lambda\over \lambda}\;
\gamma_s\left(\as(\lambda)\right)
+
\int_{\sqrt{s}/\nu}^{\sqrt{s}/\left[2(\nu/2)^{1/(2-a)}\right]}
{d\lambda\over \lambda}\;
\gamma_s\left(\as(\lambda)\right)
\nonumber\\
&\ & \hspace{-25mm} =
\int_{\sqrt{s}/\left[2(\nu/2)^{1/(2-a)}\right]}^{\sqrt{s}/2}
{d\lambda\over \lambda}\;
\gamma_s\left(\as(\lambda)\right) + (1-a)
\int_{\sqrt{s}/\left[2(\nu/2)^{1/(2-a)}\right]}^{\sqrt{s}/2}
{d\lambda\over \lambda}\;
\gamma_s\left(\as\left( \frac{s^{1-a/2}}{\nu (2 \lambda)^{1-a}}
\right)\right) \nonumber \\
&\ & \hspace{-25mm} =
(2-a)\int_{\sqrt{s}/\left[2(\nu/2)^{1/(2-a)}\right]}^{\sqrt{s}/2}
{d\lambda\over
\lambda}\;
\gamma_s\left(\as(\lambda)\right) \nonumber\\
&\ & \hspace{-15mm}
- (1-a)\int_{\sqrt{s}/\left[2(\nu/2)^{1/(2-a)}\right]}^{\sqrt{s}/2}
{d\lambda\over \lambda}\; \int_{s^{1-a/2}/\left[\nu (2 \lambda)^{1-a}
\right]}^\lambda
\frac{d \lambda'}{\lambda'}
\beta(g(\lambda'))\, {\partial \over \partial g}\,
\gamma_s\left(\as(\lambda')\right)\, .
\label{split}
\ea
In the first equality we split the $\lambda$ integral so that the limits of
the first term match those of the $B'_c$ integral of Eq.\ (\ref{globalend}).
In the second equality we have changed variables in the
second term according to
\be
\lambda \rightarrow \left({s^{1-{a\over 2}}\over
2^{1-a}\nu\lambda}\right)^{1\over 1-a}\, ,
\ee
so that the limits of the second integral also match.
In the third equality of Eq. (\ref{split}),
   we have reexpressed the running coupling at the old
scale
$\lambda$
   in terms of the new scale.
This is a generalization of the procedure of Ref.\ \cite{CT91},
applied originally to the threshold-resummed
Drell-Yan cross section \cite{DYold}.

Using Eq.\ (\ref{split}), and
identifying $p_{J_c} \cdot \xic$ with $\sqrt{s}/2$
(Eq. (\ref{xiid})) in the inclusive event shape
distribution, Eq. (\ref{globalend}),
we can rewrite this distribution at NLL as
\ba
\frac{d \sigma^{\rm incl} \left(\nu,s,a \right)}{ d \hat{n}_1 }
&=&
{d \sigma_0 \over d\hat{n}_1}\
\nonumber \\
&   & \hspace*{-27mm} \times\ \prod_{c=1}^2\,
    \exp \left\{ - \int\limits_{\sqrt{s} /
\left[2(\nu/2)^{1/(2-a)}\right]}^{\sqrt{s}/2}
\frac{d \lambda}{\lambda}  \left[ B_c
\left(c_1,c_2,a,\as\left(\lambda\right)\right)
+   2 \int\limits_{c_1 \frac{s^{1-a/2} }{ \nu
(2\,\lambda)^{1-a} } }^{c_2\, \lambda}\frac{d \lambda'}{\lambda'} A_c\left(
c_1,
a,\as\left(\lambda'\right) \right) \right] \right\}\, , \nonumber \\
    \label{globalendcat}
\ea
where we have rearranged the contribution of $\gamma_s$ as:
\ba
A_c \left( c_1, a,\as\left(\mu \right) \right) & \equiv &
A'_c \left( c_1, a,\as\left(\mu \right) \right)
- \frac{1}{4} (1-a)\,  \beta(g(\mu))\, {\partial \over \partial g}\,
\gamma_s\left(\as(\mu)\right) ,  \nonumber \\
   B_c \left(c_1,c_2,a,\as\left(\mu\right)\right) & \equiv &
\gamma_{J_c} \left(\as(\mu) \right)
   + \left( 1 - \frac{a}{2} \right) \gamma_s \left(\as(\mu) \right) +
   B'_c \left(c_1,c_2,a,\as\left(\mu\right)\right).
   \ea
Next, we replace
   the lower limit of the $\lambda'$-integral
by an explicit $\theta$-function. Then we exchange orders of integration,
and change variables in the term containing $A$
from the dimensionful variable $\lambda$ to the dimensionless combination
\be
u = {2\lambda\lambda'\over s}\, .
\ee
We find
\ba
\frac{d \sigma^{\rm incl} \left(\nu,s,a \right)}{ d \hat{n}_1 }
&=&
{d \sigma_0 \over d\hat{n}_1}\  \prod_{c=1}^2\,
    \exp \left\{ - \int\limits_{\sqrt{s} /
\left[2(\nu/2)^{1/(2-a)}\right]}^{\sqrt{s}/2}
\frac{d \lambda}{\lambda}
   B_c \left(c_1,c_2,a,\as\left(\lambda\right)\right) \right\}
\nonumber\\
&\ & \times\
\prod_{c=1}^2\,
    \exp \left\{ - 2 \int_0^{\sqrt{s}} \frac{d \lambda'}{\lambda'} \;
\int_{\lambda'{}^2/s}^{\lambda'/\sqrt{s}}\, {d u \over u}\;
\theta\left( c_1^{-1}\nu\, {\lambda'{}^a u^{1-a}\over s^{a/2}}-1
\right)\;
A_c\left( c_1,
a,\as\left(\lambda'\right) \right) \right\} \, . \nonumber \\
    \label{globalendcat2}
\ea
Here, the $\theta$-function vanishes for small $\lambda'$, and the 
remaining effects of
replacing the lower boundary of the $\lambda'$ integral by 0 are
next-to-next-to-leading logarithmic.

A further change of variables allows us to write the NLL resummed event shapes
in a form familiar from the NLL resummed thrust.
In the first line of Eq. (\ref{globalendcat2}), we
replace $\lambda^2 \rightarrow u s/4$. In the second line we relabel
$\lambda' \rightarrow \sqrt{q^2}$,
and exchange orders of integration.
Finally, choosing
\ba
c_1 & = & e^{-\gamma_E}, \nonumber \\
c_2 & = & 2,
\label{cipick}
\ea
we find at NLL
\ba
\frac{d \sigma^{\rm incl} \left(\nu,s,a \right)}{ d \hat{n}_1 }
&=&
{d \sigma_0 \over d\hat{n}_1}\  \prod_{c=1}^2\,
    \exp \left\{ \int\limits_0^1 \frac{d u}{u} \left[ \,
    \int\limits_{u^2 s}^{us} \frac{d q^2}{q^2} A_c\left(\as(q^2)\right)
    \left( e^{- u^{1-a} \nu \left(q^2/s\right)^{a/2} }-1 \right)
\right. \right.\nonumber \\
    & & \qquad \qquad
    \qquad \qquad \quad
    + \frac{1}{2} B_c\left(\as(u s/4)\right) \left( e^{-u
\left(\nu/2\right)^{2/(2-a)} e^{-\gamma_E}} -1 \right)
    \bigg] \Bigg\},
\label{thrustcomp}
\ea
and reproduce the well-known coefficients
\ba
A_c^{(1)}  & = & C_F
, \\
A_c^{(2)}  & = & \frac{1}{2} C_F  K, \\
B_c^{(1)} & = & - \frac{3}{2} \, C_F,
\ea
independent of $a$.  In Eq.\ (\ref{thrustcomp}), we have made use of the
relation
\be
e^{-x/ y} - 1 \approx - \theta \left(x - y\, e^{-\gamma_E} \right),
\ee
which is valid at NLL in the logarithmic integrals.
With these choices, when $a = 0$ we reproduce the
NLL resummed thrust cross section \cite{thrustresum}.

The choices of
the $c_i$ in Eq.\ (\ref{cipick}) cancel all purely soft NLL
components ($\gamma_s$ and $K_c$). The
remaining double logarithms stem from simultaneously soft and collinear
radiation, and single logarithms arise from collinear configurations only.
At NLL, the cross section is determined by the
anomalous dimension $A_c$, which is the coefficient
of the singular $1/[1-x]_+$ term in the
nonsinglet evolution kernel \cite{oneover1x}, and the
quark anomalous dimension.
All radiation in dijet events
thus appears to be emitted coherently by the two jets \cite{thrustresum}.
This, however, is not necessarily true
beyond next-to-leading logarithmic accuracy for dijets, and
is certainly not the case for multijet events \cite{KOS}.  Similar 
considerations apply to
the resummed correlation, Eq.\ (\ref{evolend}).

\subsection{Closed expressions}

Given the explicit results above, the integrals in  the
exponents of the resummed correlation, Eq.\ (\ref{evolend}), may
be easily performed in closed form.
We give the analytic results for the exponents
of Eq. (\ref{evolend}), as defined in
Eqs. (\ref{E1}) and  (\ref{E2}). As in Eq. (\ref{xiid}),
we identify $p_{J_c} \cdot \xic$ with
$\sqrt{s}/2$.
\ba
e^{E_1(a)} & = & \left(
\frac{ \as(\sqrt{s}/ 2)}{\as(\varepsilon \sqrt{s})}\right)^{\frac{4
C_F}{\beta_0}}
\left( \frac{\as\left(\frac{\sqrt{s}}{2 \,\zeta_0}\right)}{ \as(\sqrt{s}/2)}
   \right)^{\frac{6
C_F}{\beta_0} } , \label{E1result} \\
e^{E_2(a)} & = &   \left( \frac{ \as(c_2 \, \sqrt{s}/2)}
{\as\left(\frac{c_2 \,\sqrt{s}}{2 \, \zeta_0} \right)} \right)^{\frac{4
C_F}{\beta_0} \kappa_1(a)}
    \left( \frac{  \as\left(\frac{c_1 \, \sqrt{s}}{2 \,\zeta_0 }\right)} {
\as\left(\frac{c_1 \, \sqrt{s}}{\nu}\right)}
\right)^{\frac{1}{a-1} \frac{4 C_F}{\beta_0} \kappa_2(a)} \,
\left( \frac{ \as(c_2 \, \sqrt{s}/2)}{\as\left(\frac{c_1 \, \sqrt{s}}{2 \,
\zeta_0 }\right)} \right)^{
\frac{1}{2-a} \, \frac{8
C_F}{\beta_0} \, \ln (\nu / 2)}, \nonumber \\
& &  \label{E2result}
\ea
with
\ba
\kappa_1(a) \!\!& = &\!\! \ln \left(\frac{4}{c_2^2 e}\right) +
\frac{4\pi}{\beta_0} \left[\as\left(\frac{c_2 \, \sqrt{s}}{2 \,\zeta_0}\right)
\right]^{-1}
- \frac{2 K}{\beta_0}
- \frac{\beta_1}{2 \beta_0^2} \ln \left( \left(\frac{\beta_0}{4 \pi
e}\right)^2
\as\left(\frac{c_2\, \sqrt{s}}{2}\right) \as\left(\frac{c_2\, \sqrt{s}}{2 \,
\zeta_0}\right)\right), \nonumber \\
& & \label{C1} \\
\kappa_2(a) \!\!& = & \!\!(1 - a - 2 \gamma_E) + \frac{4\pi}{\beta_0}
\left[\as\left(\frac{\sqrt{s}}{\nu}\right)\right]^{-1}
- \frac{2 K}{\beta_0}
- \frac{\beta_1}{2 \beta_0^2} \ln \left( \left(\frac{\beta_0}{4 \pi
e}\right)^2
\as\left(\frac{c_1 \, \sqrt{s}}{\nu}\right) \as\left(\frac{c_1 \, \sqrt{s}}{2
\, \zeta_0}\right)\right). \nonumber \\
& & \label{C2}
\ea
   We have used the two-loop running coupling, when appropriate,
to derive Eqs.\ (\ref{E1result}) - (\ref{C2}).
The results are expressed in terms of the one-loop running coupling
\be
\as(\mu) = \frac{2 \pi}{\beta_0} \frac{1}{\ln \left(
\frac{\mu}{\Lambda_{\mbox{\tiny QCD} } }  \right)}\, ,
\ee
and the first two coefficients in the expansion of the QCD beta-function,
$\beta_0$ and
\be
\beta_1 = \frac{34}{3} \, C_A^2 - \left(\frac{20}{3} C_A  + 4 C_F\right) \,
T_F \, N_f\, .
\ee
Combining the expressions for the exponents, Eqs.\ (\ref{E1result}) and
(\ref{E2result}), for the Born cross section,
Eq.\ (\ref{bornCross}), and for the soft function, Eq.\ (\ref{oneLoopSoft}), in
Eq.\ (\ref{evolend}),
the complete differential cross section, at LL in $\varepsilon$ and at NLL
in $\nu$, is given by
\ba
\frac{d \sigma \left(\varepsilon, \nu,s ,a\right)}{d\varepsilon\, d
\hat{n}_1 }
&=&
\Ncol \left( \sum_{\rm f} Q_{\rm f}^2 \right) \frac{\pi \alpha_{\mbox{\tiny
em}}^2}{2 s} \left(1+ \cos^2 \theta \right) C_F \frac{\as(\varepsilon
\sqrt{s})}{\pi} \frac{1}{\varepsilon} \int\limits_\O  d \mbox{PS}_2\,
\frac{1}{2 \pi} \frac{\beta_1 \cdot \beta_2}{\beta_1
\cdot \hat{k} \, \beta_2 \cdot \hat{k} } \nonumber \\
& \times &
\left(
\frac{ \as\left(\frac{\sqrt{s}}{2}\right)}{\as(\varepsilon
\sqrt{s})}\right)^{\frac{4 C_F}{\beta_0}}
\left( \frac{\as\left(\frac{\sqrt{s}}{2 \,\zeta_0}\right)}{
\as\left(\frac{\sqrt{s}}{2}\right)}
   \right)^{\frac{6
C_F}{\beta_0} }
   \nonumber \\
&  \times & \left( \frac{ \as\left(c_2 \, \frac{\sqrt{s}}{2}\right)}
{\as\left(\frac{c_2 \,\sqrt{s}}{2 \, \zeta_0} \right)} \right)^{\frac{4
C_F}{\beta_0} \kappa_1(a)}
    \left( \frac{  \as\left(\frac{c_1 \, \sqrt{s}}{2 \,\zeta_0 }\right)} {
\as\left(\frac{c_1 \, \sqrt{s}}{\nu}\right)}
\right)^{\frac{1}{a-1} \frac{4 C_F}{\beta_0} \kappa_2(a)} \,
\left( \frac{ \as\left(c_2 \, \frac{\sqrt{s}}{2}\right)}{\as\left(\frac{c_1
\, \sqrt{s}}{2 \, \zeta_0 }\right)} \right)^{
\frac{1}{2-a} \, \frac{8
C_F}{\beta_0} \, \ln \left(\frac{\nu}{2}\right)}\hspace*{-2.0cm} .
\label{resulto1}
\ea
These are the expressions that we will evaluate in the
next section.  We note that this is not the only possible closed form for the
resummed correlation at this level of accuracy.  When a full
next-to-leading order calculation for this set of event shapes
is given, the matching procedure of
\cite{thrustresum} may be more convenient.

\section{Numerical Results}
\label{numerics}

Here we show some representative examples of numerical results for the
correlation, Eq.\ (\ref{resulto1}). We pick the constants $c_i$ as in
Eq.\ (\ref{cipick}),
unless stated otherwise. The effect of different choices
is nonleading, and is numerically small, as we will see below.
In the following we choose the
region $\O$ to be a ring between the jets, centered in their
center-of-mass, with a width of $\Delta \eta = 2$, or equivalently, opening
angle $\delta \approx 50$ degrees (see Eq.\ (\ref{rapidity})).
The analogous cross section for
   a cone centered at 90 degrees from the jets
(Eq.\ (\ref{softcone})) has a similar behavior.
In the following, the center-of-mass energy
$Q=\sqrt{s}$ is  chosen to be $100$ GeV.

\begin{figure}[htb]
\vspace*{7mm}
\begin{center}
a) \hspace*{7.5cm} b) \hspace*{4cm} \vspace*{-6mm} \\
\epsfig{file=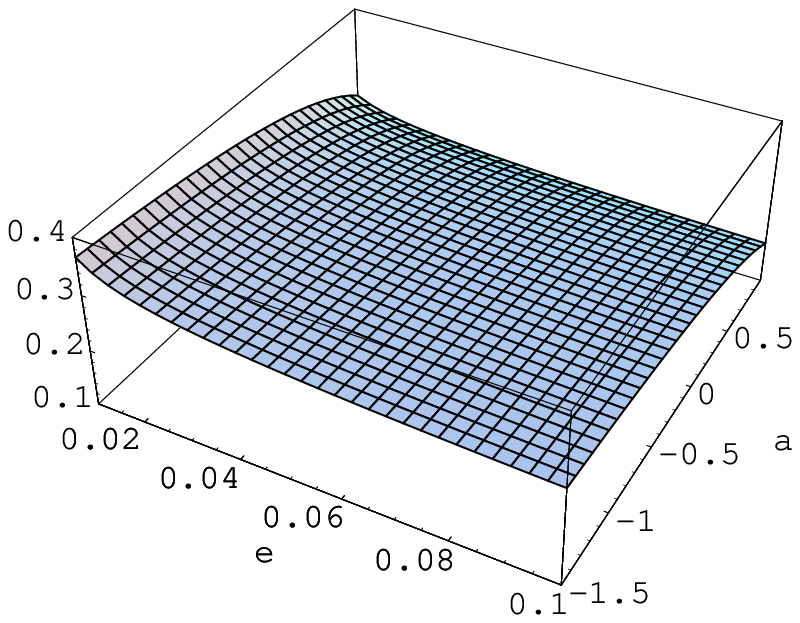,width=6.8cm,clip=0}  \mbox{
}\hspace*{5mm}  \epsfig{file=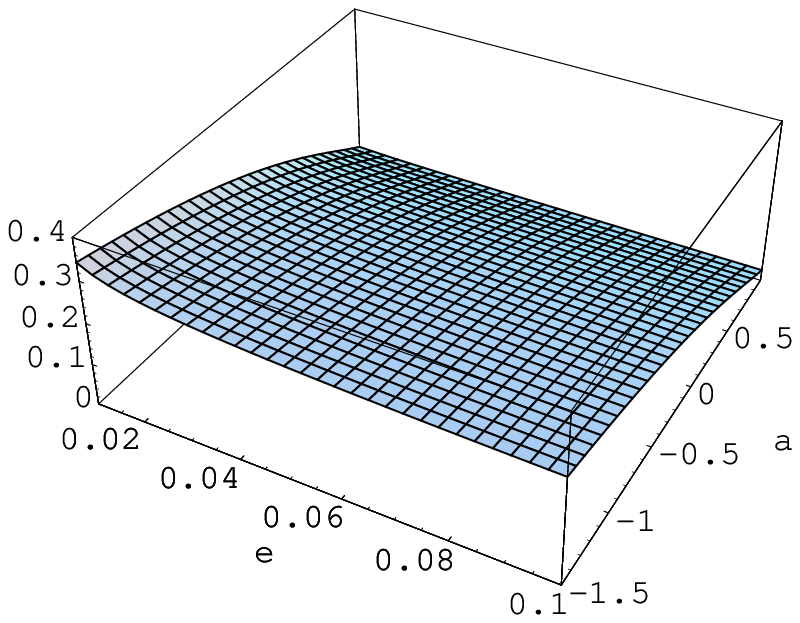,width=6.8cm,clip=0}
\caption{Differential cross section $\frac{\varepsilon d
\sigma/(d\varepsilon d\hat n_1)}{d \sigma_0/d \hat n_1}$,
normalized by the Born cross section, at $Q = 100$ GeV,
as a function of
$\varepsilon$ and $a$ at fixed $\nu$: a) $\nu = 10$, b) $\nu =
50$. $\O$ is a ring (slice) centered around the jets, with a
width of $\Delta \eta = 2$.} \label{num1}
\end{center}
\end{figure}

Fig. \ref{num1} shows the dependence of the differential cross
section (\ref{evolend}), multiplied by $\varepsilon$
and normalized by the Born cross section,
$\frac{\varepsilon d \sigma/(d\varepsilon d\hat n_1)}{d
\sigma_0/d \hat n_1}$, on the measured energy $\varepsilon$ and
on the parameter $a$, at fixed $\nu$.
In Fig. \ref{num1} a), we plot $\frac{\varepsilon
d \sigma/(d\varepsilon d\hat n_1)}{d \sigma_0/d \hat n_1}$ for
$\nu = 10$, in Fig. \ref{num1} b) for $\nu  = 50$.
As $\nu$ increases, the radiation into the
complementary region $\bar \O$ is more restricted,
as illustrated by the comparison of Figs.
\ref{num1} a)  and b). Similarly, as $a$ approaches 1, the
cross section falls, because the jets are restricted to
be very narrow.  On the other hand,
as $a$ assumes more and more negative values at
fixed $\varepsilon$, the correlations (\ref{evolend}) approach a
constant value.  For $a$ large and negative, however, non-global
dependence on
$\ln\varepsilon$ and $|a|$ will emerge from higher order corrections
in the soft function, which we do not include in Eq.\ (\ref{resulto1}).

\begin{figure}[htb]
\begin{center}
\epsfig{file=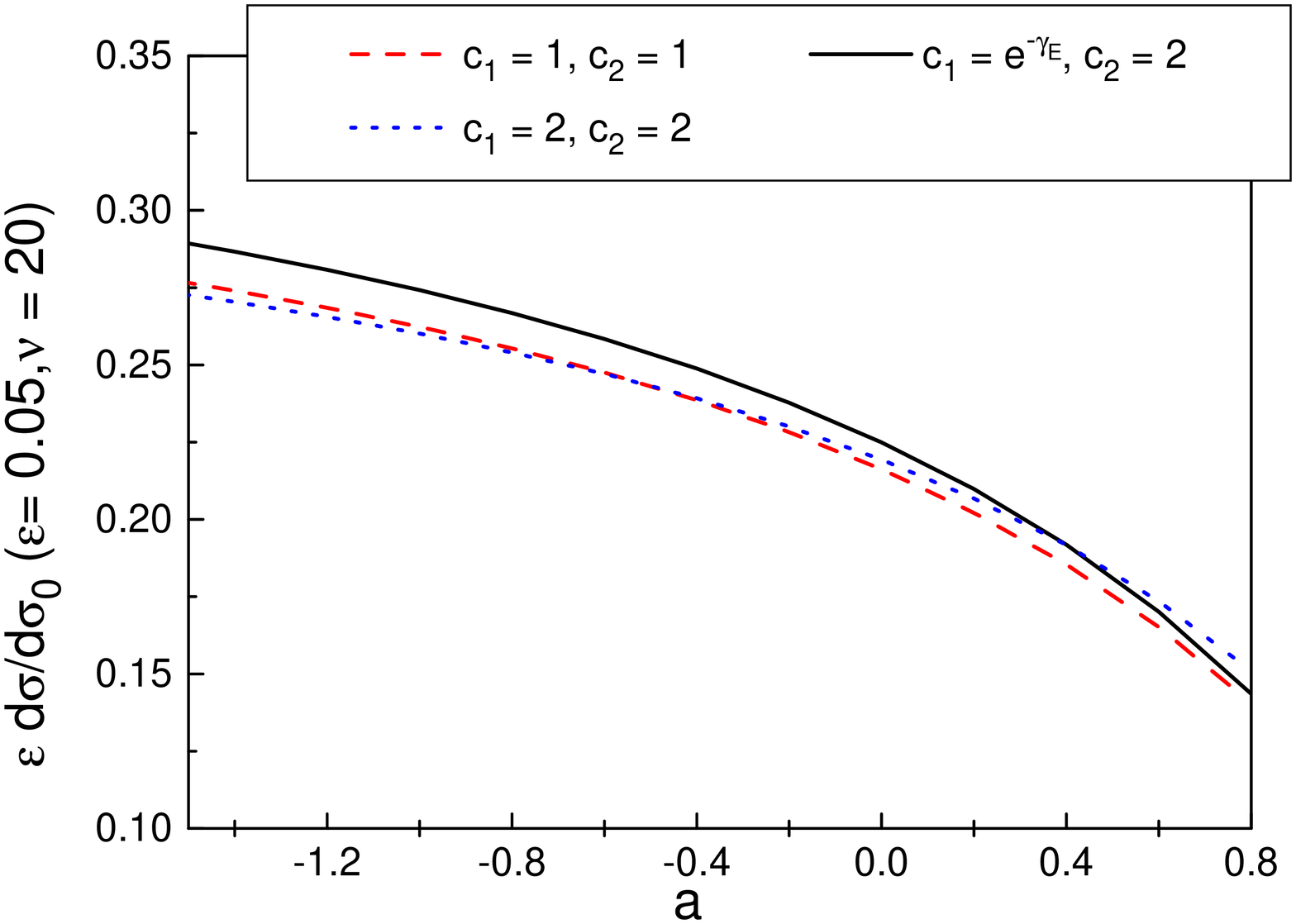,height=10cm,angle=270,clip=0}
\caption{Differential cross section $\frac{\varepsilon d
\sigma/(d\varepsilon d\hat n_1)}{d \sigma_0/d \hat n_1}$,
normalized by the Born cross section, at $Q = 100$ GeV,
as a function of
$a$ at fixed $\nu = 20$ and $\varepsilon = 0.05$.
   $\O$ is chosen as in Fig. \ref{num1}. Solid line: $c_1 = e^{-\gamma_E},\,
c_2 = 2$,
as in Eq.  (\ref{cipick}), dashed line: $c_1 = c_2 = 1$, dotted line:
$c_1 = c_2 = 2$. } \label{numci}
\end{center}
\end{figure}

In Fig.\ \ref{numci} we investigate the sensitivity of the resummed
correlation,
Eq.\ (\ref{resulto1}), to our choice of the constants $c_i$. The 
effect of these constants is
of next-to-next-to-leading logarithmic order in the event shape. We plot the
differential cross section $\varepsilon
\frac{\varepsilon d \sigma/(d\varepsilon d\hat n_1)}{d
\sigma_0/d \hat n_1}$, at $Q = 100$ GeV, for fixed $\varepsilon = 0.05$ and
$\nu = 20$, as a function of $a$. The effects of changes in the $c_i$ are
of the order of a few percent for moderate values of $a$.

\begin{figure}[htb]
\vspace*{6mm}
\begin{center}
a) \hspace*{7.5cm} b) \hspace*{4cm} \vspace*{-6mm} \\
\epsfig{file=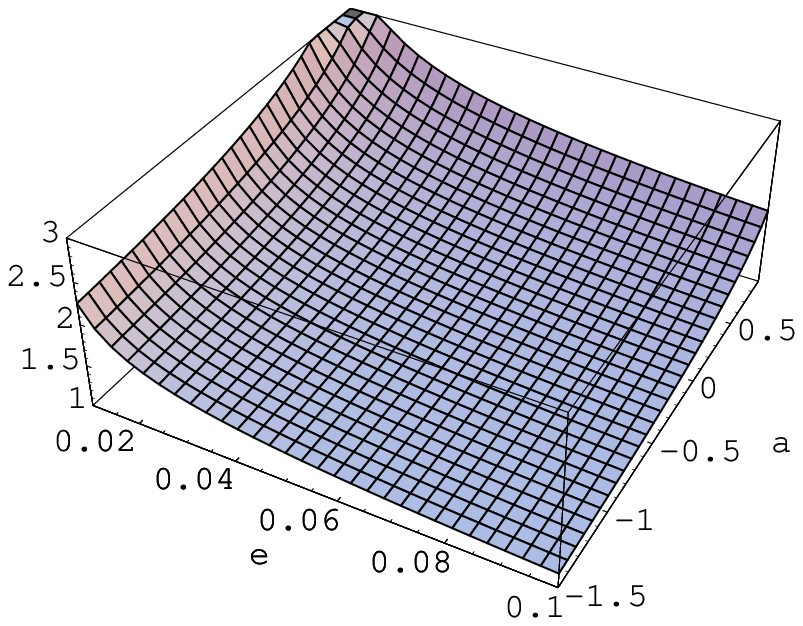,width=6.7cm,clip=0}  \mbox{
}\hspace*{5mm}  \epsfig{file=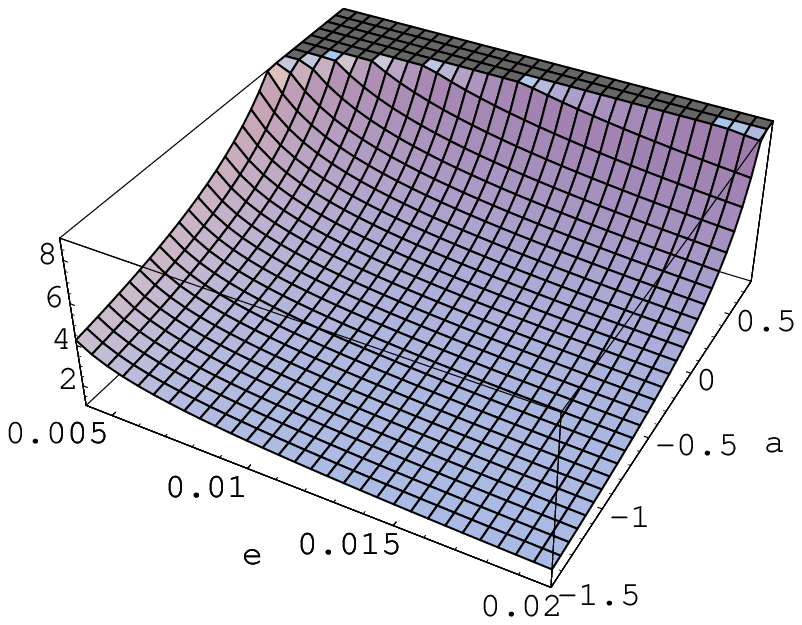,width=6.7cm,clip=0}
\caption{Ratios of differential cross sections  for quark to
gluon jets $\frac{C_A}{C_F} \left(\frac{\varepsilon d
\sigma^q/(d\varepsilon d\hat n_1)}{d \sigma_0^q/d \hat n_1}\right)
\left(\frac{\varepsilon d
\sigma^g/(d\varepsilon d\hat n_1)}{d \sigma_0^g/d \hat n_1}\right)^{-1}$ at
$Q = 100$ GeV
as a
function of $\varepsilon$ and $a$ at fixed $\nu$: a) $\nu = 10$,
b) $\nu = 50$. $\O$ as in Fig. \ref{num1}, $c_1$ and $c_2$ as in Eq.
(\ref{cipick}).}
\label{num2}
\end{center}
\end{figure}

Finally, we illustrate the sensitivity of these results to the flavor
of
the primary partons. For this purpose we study the corresponding 
ratio of the shape/flow
correlation to the cross section for gluon jets produced
by a hypothetical color singlet source. Fig.\ \ref{num2} displays the ratio
of the differential cross section
$d\sigma^q(\varepsilon,a)/(d\varepsilon d \hat n_1)$, Eq.\ (\ref{resulto1}),
normalized
by the lowest-order cross section, to the
analogous quantity with gluons as
primary partons in the outgoing jets, again at
$Q = 100$ GeV.
This ratio is multiplied by $C_A/C_F$ in the figure to compensate for
the difference in the normalizations of the lowest-order soft functions.
Gluon jets have wider angular extent, and hence are
suppressed relative to quark jets with increasing  $\nu$ or $a$,
as can be seen by
comparing Figs. \ref{num2} a) and b). Fig. \ref{num2} a) shows the ratio
at $\nu = 10$, and Fig. \ref{num2} b) at $\nu = 50$.
These results suggest sensitivity to
the more complex color and flavor flow
characteristic of hadronic scattering \cite{KOS,BKS1}.

\section{Summary and Outlook}

We have
introduced a general class of inclusive event shapes in $e^+e^-$ dijet events
which
reduce to the thrust and the jet broadening distributions as special
cases.
We have derived analytic expressions in transform
space, and have shown the equivalence of our formalism at NLL
with the well-known result for the thrust \cite{thrustresum}.
Separate studies of this class of event shapes
in the untransformed space,
at higher orders, and for nonperturbative effects
\cite{ptnpdist} are certainly of interest.
We reserve these studies
for future work.

We have introduced a set of correlations of interjet energy
flow for the general class of  event shapes, and have shown
that for these quantities it is possible to control
the influence of secondary radiation and nonglobal logarithms.
These correlations are sensitive mainly to radiation
emitted directly from the primary hard scattering, through transforms in
the weight functions that suppress
secondary, or non-global, radiation.
  We have presented analytic  and
numerical studies of these shape/flow correlations at leading
logarithmic order in the
flow variable and at next-to-leading-logarithmic order in the
event shape.
   The application of our formalism
to multijet events and to scattering with initial state hadrons is
certainly possible, and may shed light on the relationship
between color and energy flow in hard scattering processes with non-trivial
color exchange.

\subsection*{Acknowledgements}

We thank Maria Elena Tejeda-Yeomans for many helpful discussions,
and Gavin Salam for useful conversations.
   This work was supported in part by
the National Science Foundation grant PHY0098527.

\begin{appendix}

\section{Eikonal Example}
\label{eikapp}

In this appendix, we give details of the
calculation of the logarithmic behavior in the
diagrams of Fig.\ \ref{diagrams}.
We choose the reference frame such
that the momenta of the final
state particles are given by:
\ba \label{momenta}
{\beta}_1 & = & (1,0,0, 1), \nonumber \\
{\beta}_2 & = & (1,0,0,-1), \nonumber \\
l & = & \ol (1, s_l, 0, c_l), \nonumber \\
k & = & \ok (1, s_k \cfi, s_k \sfi, c_k).
\ea
Here we define $s_{l,k} \equiv \sin
{\theta}_{l,k}$ and $c_{l,k} \equiv
\cos {\theta}_{l,k}$. $\theta_l$ is the angle between the vectors
$\vec{l}$ and $\vec{\beta_1}$,  $\theta_k$ is the
angle between the vectors $\vec{k}$ and $\vec{\beta_1}$
and $\phi$ is the azimuthal angle
of the gluon with momentum $k$ relative to the plane defined by
$\beta_1$, $\beta_2$ and $l$. The
available phase space in polar angle for the radiated gluons is
${\theta}_k \in (\pi/2 - \delta, \,
\pi/2 + \delta)$ and ${\theta}_l \in (0,\, \pi/2 - \delta) \cup
(\pi/2 + \delta, \, \pi)$.

Using the diagrammatic rules for eikonal lines and
vertices, as listed for example in \cite{pQCD}, we can write down the
expressions corresponding to each diagram separately. For example,
diagram \ref{diagrams} a) gives
\ba \label{a}
a) \, + \, (k \leftrightarrow l)\!\! &\!\! =\!\! &\!\! \left[ f_{abc} 
\mathrm{Tr}(t_a
t_b t_c) \right] \left( -i g_s^4 \,
\beta_1^{\alpha} \beta_2^{\beta} \beta_1^{\gamma} \right) \,
V_{\alpha \beta \gamma}(k+l, -k, -l) \, \frac{1}{\beta_1
\cdot (k+l)} \, \frac{1} {2 k \cdot l} \, \frac{1}{\beta_1 \cdot l}
\, \frac{1}{\beta_2 \cdot k} \nonumber \\
& + &  \, (k \leftrightarrow l).
\eea
$V_{\alpha \beta \gamma}(k+l, -k, -l) =
[ (2k+l)_{\gamma} g_{\alpha \beta} + (l-k)_{\alpha} g_{\beta \gamma}
- (2l + k)_{\beta} g_{\alpha
\gamma}]$ is the
momentum-dependent part of the three gluon vertex.
Using the color identity $f_{abc} \mathrm{Tr}(t_a t_b t_c) = i C_F \Ncol
C_A /2$, and the approximation $\beta_j \cdot l \gg \beta_j \cdot
k$ for $j=1,2$, which is valid due
to the strong ordering of the final state gluon energies, we arrive at
\be
a) + (k \leftrightarrow l) = \frac{1}{4} \, C_F \Ncol C_A \, g_s^4 \,
\frac{\beta_1 \cdot \beta_2}{k
\cdot l}
\left( \frac{1}{\beta_1 \cdot k \, \beta_2 \cdot l} +
\frac{2}{\beta_1 \cdot l \, \beta_2 \cdot k}
\right).
\ee
We proceed in a similar manner for the rest of the diagrams. The
results are:
\ba \label{b-e}
b) + (k \leftrightarrow l) & = & \frac{1}{4} \, C_F \Ncol C_A \, g_s^4 \,
\frac{\beta_1
\cdot \beta_2}{k \cdot l}\left(
\frac{2}{\beta_1 \cdot k \, \beta_2 \cdot l} + \frac{1}{\beta_1 \cdot
l \, \beta_2
\cdot k}\right), \nonumber \\
c) & = & \frac{1}{4} \, C_F \Ncol C_A \, g_s^4 \, \frac{\beta_1 \cdot \beta_2}
{k \cdot l}\frac{1}{\beta_1 \cdot l} \frac{1}{\beta_2 \cdot k}, \nonumber
\\
d) & = & \frac{1}{4} \, C_F \Ncol C_A \, g_s^4 \, \frac{\beta_1 \cdot \beta_2}
{k \cdot l}\frac{1}{\beta_1 \cdot k} \frac{1}{\beta_2 \cdot l}, \nonumber
\\
e) & = & C_F \Ncol (C_F - C_A/2) \, g_s^4 \, \frac{(\beta_1 \cdot \beta_2)^2}
{\beta_1 \cdot l \, \beta_2 \cdot l} \frac{1}{\beta_1 \cdot k \,
\beta_2 \cdot k}, \nonumber \\
f) + (k \leftrightarrow l) & = & C_F \Ncol (C_F - C_A/2) \, g_s^4 \,
\frac{(\beta_1 \cdot \beta_2)^2}{\beta_1 \cdot l \, \beta_2 \cdot l}
\frac{2}{\beta_1 \cdot k \, \beta_2 \cdot k}.
\ea
The color factors in the last two equations of
(\ref{b-e})
are obtained from the identity $\mathrm{Tr}(t_a t_b t_a t_b) = C_F \Ncol
(C_F - C_A/2)$. Combining the terms proportional to
the color factor $C_F \Ncol C_A$, and including the complex conjugate
diagrams, we find for the squared amplitude
\be \label{mm}
|M|^2 = 2 \, g_s^4 \, C_F \Ncol C_A \, \beta_1 \cdot \beta_2 \left(
\frac{1}{k \cdot \l \, \beta _1
\cdot k \, \beta_2 \cdot l} +
\frac{1}{k \cdot \l \, \beta _1 \cdot l \, \beta_2 \cdot k} -
\frac{\beta_1 \cdot \beta_2}{\beta_1
\cdot l \, \beta_2 \cdot l \, \beta_1 \cdot k \, \beta_2 \cdot k} \right).
\ee
Having determined the amplitude, we need to integrate $|M|^2$ over
the phase space corresponding
to the geometry given in Fig. \ref{kinematics}. Specifically, we have
to evaluate:
\be \label{ps}
I \equiv \frac{1}{\Ncol} \int {\mathrm d}{\bar \varepsilon} \, e^{-\nu \,
{\bar \varepsilon}}
\, \int_{\Omega}
\frac{{\mathrm d}^3 k}{(2\pi)^3 \, 2 \omega_k} \,
\int_{\bar{\Omega}} \frac{{\mathrm d}^3 l}{(2\pi)^3 \, 2 \omega_l} \,
\delta(\varepsilon - \omega_k
/ \sqrt{s}) \, \delta({\bar \varepsilon} - {\bar f}(l,a))
\, |M|^2,
\ee
where the weight function ${\bar f}(l,a)$ is given, as in Eqs.\
(\ref{2jetf})
and (\ref{fbarexp}), by
\be \label{fb}
{\bar f}(l,a) = \left\{ \begin{array}{l@{\quad:\quad}l}
\frac{\omega_l}{\sqrt{s}} \, (1-c_l)^{1-a} \, s_l^a & \theta_l \in
(0, \, \pi/2 - \delta) \\
\frac{\omega_l}{\sqrt{s}} \, (1+c_l)^{1-a} \, s_l^a & \theta_l \in
(\pi/2 + \delta, \, \pi),
\end{array} \right.
\ee
with $a < 1$.

\begin{figure} \center
\epsfig{file=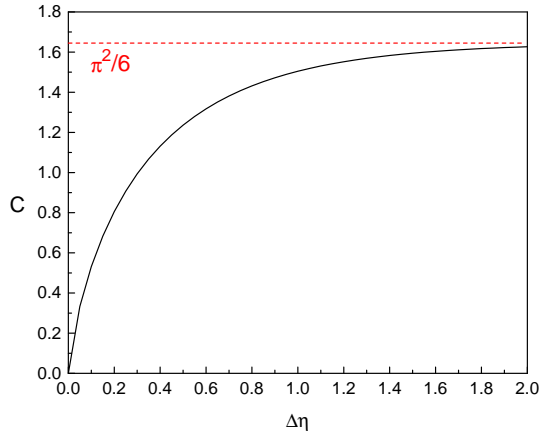,height=8cm,angle=270,clip=0}
\caption{\label{crossSec_vs_rapidity} $C(\Delta \eta)$, as defined in 
(\ref{Cdef}),
as a function of rapidity
width
$\de$ of the region $\Omega$. The dashed line is its limiting value,
$C (\de \rightarrow \infty) = \pi^2/6$.}
\end{figure}

Using the equalities: $\beta_1 \cdot \beta_2 = 2$, $\beta_1 \cdot l =
\ol(1-c_l)$,
$\beta_2 \cdot l = \ol(1 + c_l)$, $\beta_1 \cdot k = \ok(1 - c_k)$,
$\beta_2 \cdot k = \ok(1 + c_k)$ and $k \cdot l =
\ok \ol (1 - c_k c_l - s_k s_l \cfi)$ in Eq. (\ref{mm}), performing
the integration
   over $\phi$, and changing the integration variable
$c_l \rightarrow - c_l$ in the angular region $\theta_l \in (\pi/2 +
\delta, \, \pi)$, we easily arrive at the following
three-dimensional integral:
\bea \label{ps1}
I & = & C_F C_A \left(\frac{\alpha_s}{\pi}\right)^2 \,
\frac{1}{\varepsilon} \,
\int_{-\sd}^{\sd}
\mathrm{d} c_k \, \int_{\sd}^{1}
\mathrm{d} c_l \, \int_{\varepsilon \sqrt{s}}^{\sqrt{s}} \frac{\mathrm{d}
\ol}{\ol} \, e^{-\nu \, \ol \,
(1-c_l)^{1-a} \, s_l^a / \sqrt{s}}
\nonumber \\
& & \left[ \frac{1}{c_k + c_l} \, \frac{1}{1+c_k}
\left(\frac{1}{1+c_l} + \frac{1}{1-c_k}\right) - \frac{1}{s_k^2} \,
\frac{1}{1+c_l} \right].
\eea
We are interested in the $(1/\varepsilon)\ln(1/\varepsilon)$ behavior of $I$.
This is
obtained after performing the $\ol$ integral with
the replacement $e^{-\nu \ol (1-c_l)^{1-a} \, s_l^a  / \sqrt{s}} \rightarrow
\theta(1 - \nu \ol (1-c_l)^{1-a} \, s_l^a  /
\sqrt{s})$. Remainders do
not contain terms proportional to $\ln \varepsilon$.
In this approximation, the $c_l$ integration can be carried out, and
we obtain the integral representation for the
term containing $(1/\varepsilon)\ln(1/\varepsilon)$:
\ba \label{ps2}
I & = & 2 \, C_F C_A \left(\frac{\alpha_s}{\pi}\right)^2 \,
\frac{1}{\varepsilon} \,
\ln\left(\frac{1}{\varepsilon \nu}\right) \, \left[ \int_{0}^{\sd} \,
\frac{\mathrm{d} c_k}{s_k^2} \, \ln \left(\frac{s_k^2}{s_k^2 -
\cos^2\delta} \right) -
\ln \left(\frac{2}{1 + \sd} \right) \,
\ln \left(\frac{1+\sd}{1-\sd} \right) \right]. \nonumber \\
& &
\ea
The  potential non-global logarithm of $\varepsilon$ is replaced by 
$\ln(\varepsilon \nu)$.
The angular integral over $c_k$ can be expressed in terms of
dilogarithmic functions. The final expression for the term proportional to
$\ln(\varepsilon \nu)/\varepsilon$ takes the form:
\ba \label{ps3}
I & = & C_F C_A \left(\frac{\alpha_s}{\pi}\right)^2 \,
\frac{1}{\varepsilon} \,
\ln\left(\frac{1}{\varepsilon \nu}\right) \,
    \left[\frac{\pi^2}{6} +
\ln \left(\frac{\cot\delta \, (1 + \sd)}{4} \right) \, \ln
\left(\frac{1+\sd}{1-\sd} \right)
+ \mathrm{Li}_2\left(\frac{1-\sd}{2}\right) \right. \nonumber \\
& - & \left. \mathrm {Li}_2\left(\frac{1+\sd}{2}\right) -
\mathrm{Li}_2\left(-\frac{2 \sd}{1-\sd}\right) -
\mathrm{Li}_2\left(\frac{1-\sd}{1+\sd}\right) \right].
\ea
Equivalently, we can express our results in terms of the rapidity
width of the region $\Omega$, Eq. (\ref{rapidity}),
and we obtain
\ba \label{psRapidity}
I & = & C_F C_A \left(\frac{\alpha_s}{\pi}\right)^2 \,
\frac{1}{\varepsilon} \,
\ln\left(\frac{1}{\varepsilon \nu}\right) \,
\left[\frac{\pi^2}{6} +
\de \left(\frac{\de}{2} - \ln \left(2 \sinh (\de)\right) \right)
+ \mathrm{Li}_2 \left ( \frac{e^{-\de/2}}{2 \, \cosh (\de/2)} \right)
\right. \nonumber \\
& - & \left. \mathrm {Li}_2 \left( \frac{e^{\de/2}}{2 \, \cosh
(\de/2)} \right) - \mathrm{Li}_2
\left( -2 \sinh (\de/2) \,
e^{\de/2} \right) - \mathrm{Li}_2 (e^{-\de}) \right].
\ea
The coefficient
\be
C (\Delta \eta) \equiv - \left( \frac{\pi}{ \alpha_s} \right)^2 \, 
\frac{\varepsilon \,
I}{C_F C_A \ln (\varepsilon
\nu)} \label{Cdef}
\ee
as a
function of $\de$ is shown in
Fig. \ref{crossSec_vs_rapidity}.
Naturally, $C$ is a monotonically increasing function of $\Delta\eta$.
For $\de \rightarrow 0$,
\be
C \sim {\mathcal O}(\de \, \ln \de)\, ,
\ee
and the cross section vanishes, as expected. On the other hand, as the size of
region $\Omega$ increases, $C$ rapidly saturates and reaches its limiting
value \cite{DS}
\be
\lim_{\de \rightarrow \infty} \, C = \frac{\pi^2}{6}\, .
\ee

\section{Recoil} \label{approxapp}

In this appendix, we return to the justification
of the technical step represented by Eq.\ (\ref{nnJone}).
According to this approximation, we may
compute the jet functions by identifying
axes that depend only upon particles in the
final states
$N_{J_c}$ associated with those functions, rather
than the full final state $N$.
Intuitively, this is a reasonable estimate, given
that the jet axis should be determined by
a set of energetic, nearly collinear particles.
When we make this replacement, however,  the contributions
to the event shape from energetic particles near the jet axis may
change.  This change is neglected in going from
the original factorization, Eq.\ (\ref{sigmafact}), to the
factorization in convolution form, Eq.\ (\ref{factor}),
which is the starting point for the resummation
techniques that we employ in this paper.
The weight functions $\bar f^N(N_i,a)$
in Eq.\ (\ref{sigmafact}) are defined
relative to the unit vector $\hat n_1$ corresponding to
$a=0$, the thrust-like event shape.
The factorization of Eq.\ (\ref{sigmafact})
applies to any $a<2$, but as indicated by the superscript,
individual contributions to $\bar f^N(N_i,a)$ on the
right-hand side continue to depend on the full
final state $N$, through the identification of the jet axis.

To derive the factorization
of Eq.\ (\ref{factor}) in a simple convolution
form, we must be able to
treat the thrust axis, $\hat n_1$, as a fixed vector
for each of the states $N_s$, $N_{J_c}$.  This is possible
if we can neglect the effects of recoil from soft,
wide-angle radiation on the direction of the axis.
Specifically, we must be able to make the replacement
\be
\bar f_{\bar{\O}_c}^N(N_{J_c},a) \rightarrow \bar f_c(N_{J_c},a)\, , 
\label{replace}
\ee
where $\bar f_c(N_{J_c},a)$ is the event shape variable
for jet $c$, in which the axis $\hat n_c$ is
specified by state $N_{J_c}$ {\it only}.  Of course, this
replacement changes the value of the weight, $\bar\varepsilon$,
$\bar f_{\bar{\O}_c}^N(N_{J_c},a) \ne \bar f_c(N_{J_c},a)$.
As we now show, the error induced by this
replacement is suppressed by a power
of $\bar \varepsilon$ so long as $a<1$.  In general,
the error is nonnegligible for $a\ge 1$.
The importance of recoil for jet broadening, at
$a=1$, was pointed out in \cite{broaden2}.  We now
discuss how the neglect of such radiation
affects the jet axis
(always determined from $a=0$)
and hence the value of the event shape for arbitrary $a<2$.

The jet axis is
found by minimizing $\bar f(a=0)$
in each state.
The largest influence on the axis ${\hat n}_c$ for jet $c$
is, of course, the set of fast, collinear particles
within the state $N_{J_c}$ associated with the jet function
in Eq.\ (\ref{sigmafact}).
Soft, wide-angle radiation, however,
does affect the precise direction
of the axis.  This is what we mean by `recoil'.

Let us denote by $\o_s$ the energy of the soft wide-angle radiation that is
neglected in the factorization
(\ref{factor}).  Neglecting this soft radiation in
the determination of the jet axis
will result in an axis $\hat n_1(N_{J_c})$, which differs from the
axis $\hat n_1 (N)$
determined from the complete final state $(N)$ by an angle $\Delta_s\phi$:
\be
     \angle\!\!\!) \left(\hat n_1(N), \hat n_1(N_{J_c}) \right) \equiv
\Delta_s \phi
\sim  {\o_s \over Q}\, .
\label{deltaphi}
\ee
At the same time, the soft, wide-angle radiation also contributes
to the total event shape
$\bar f(N,a) \sim (1/Q)k_\perp^a (k^-)^{1-a}$ at the
level of
\be
\bar\varepsilon_s \sim {\o_s\over Q} \, ,
\ee
because for such wide-angle radiation, we may take $k_s^-\sim 
k_{s,\perp}\sim\omega_s$.
In summary, the neglect of wide-angle soft radiation rotates the jet axis
by an angle that is of the order of the contribution
of the same soft radiation to the event shape.

In the factorization (\ref{factor}), the contribution of
each final-state particle is taken into account,
just as in Eq.\ (\ref{sigmafact}).  The question
we must answer is how the rotation of the jet axis affects
these contributions, and hence the value of the event
shape.

For a wide-angle particle, the rotation of the jet
axis by an angle of order $\Delta_s\phi$
in Eq.\ (\ref{deltaphi}) leads to a
negligible change in its contributions to the event shape, because
its angle to the axis is a number of order unity, and the
jet axis is rotated only
by an angle of order $\bar\varepsilon_s$.
Contributions from soft radiation are therefore stable
under the approximation (\ref{nnJone}).
The only source of large corrections is then associated
with energetic jet radiation,
because these particles are nearly collinear to the jet axis.

It is easy to see from the form
of the shape function in terms of angles, Eq.\ (\ref{fbarexp}),
that for any value of parameter $a$, a particle
of energy $\omega_i$ at a small angle $\theta_i$ to
the jet axis $\hat n_1 (N)$ contributes to the
event shape at the level
\be
\bar\varepsilon_i \sim  {\omega_i \over Q}\theta_i{}^{2-a}\, .
\ee
The rotation of the jet
axis by the angle $\Delta_s\phi$ due to neglect of soft radiation
may be as large as, or larger than,
$\theta_i$. Assuming the latter, we find a shift in the
$\bar\varepsilon_i$ of order
\be
\delta \bar\varepsilon_i \equiv \bar\varepsilon_i \left( \hat n_1(N) \right)
- \bar\varepsilon_i \left( \hat n_1(N_{J_c}) \right)
\sim {\omega_i\over Q}\,
\left(\Delta_s\phi\right)^{2-a} \sim
{\omega_i\over Q}\, \left({\o_s\over Q}\right)^{2-a} \sim
{\omega_i\over Q}\, \bar\varepsilon_s{}^{2-a}\, .
\ee
The change in $\bar\varepsilon_i$
   is thus suppressed by at least a factor $\bar\varepsilon_s{}^{1-a}$
compared to $\bar\varepsilon_s$, which is the
contribution of the wide-angle soft radiation
to the event shape.  The contributions of nearly-collinear,
energetic radiation to the event shape thus change
significantly under the replacement (\ref{nnJone}),
but so long as $a<1$, these contributions are
power-suppressed in the value
of the event shape, both before and after the approximation
that leads to a rotation of the axis.
For this reason, when $a<1$ (and only when $a<1$), the value
of the event shape is stable whether or not
we include soft radiation in the determination
of the jet axes, up to corrections that are suppressed
by a power of the event shape.  In this case, the
transition from Eq.\ (\ref{sigmafact}) to Eq.\ (\ref{factor})
is justified.

\end{appendix}

\end{document}